\documentclass[aps,twocolumn,10pt,superscriptaddress,prb]{revtex4-1}

\pdfoutput=1

\usepackage{hyperref}
\usepackage{mathtools}
\usepackage{subfigure}
\usepackage{amsmath}
\usepackage{amssymb}
\usepackage{physics}
\DeclareMathAlphabet{\pazocal}{OMS}{zplm}{m}{n}
\usepackage{leftidx}
\let\counterwithin\relax
\usepackage{chngcntr}
\usepackage{color}

\newcommand{\f}[2]{\frac{#1}{#2}} 
\newcommand{\beq}{\begin{equation}}
\newcommand{\eeq}{\end{equation}}

\newcommand{\bo}[1]{\boldsymbol{#1}}

\newcommand{\D}{\mathrm{d}}

\begin{document}

\title{Weak-coupling superconductivity in an anisotropic three-dimensional repulsive Hubbard model}

\author{Henrik Schou R{\o}ising}
\email[]{henrik.roising@physics.ox.ac.uk}
\affiliation{Rudolf Peierls Center for Theoretical Physics, Oxford OX1 3PU, United Kingdom}

\author{Felix Flicker}
\affiliation{Rudolf Peierls Center for Theoretical Physics, Oxford OX1 3PU, United Kingdom}

\author{Thomas Scaffidi}
\affiliation{Department of Physics, University of California, Berkeley, California 94720, USA}

\author{Steven H. Simon}
\affiliation{Rudolf Peierls Center for Theoretical Physics, Oxford OX1 3PU, United Kingdom}

\date{\today}

\begin{abstract}
We study a three-dimensional single-band repulsive Hubbard model at weak coupling. We establish the superconducting phase diagram in the parameter space of the chemical potential and the out-of-plane hopping strength. The model continuously connects the Hubbard model in two and three dimensions. We confirm previously-established results in these limits, and identify a rich structure of competing order parameters in between. Specifically, we find five types of $p$- and $d$-wave orders. In several regions of the phase diagram, even when the Fermi surface is a corrugated cylinder, the ground state is a time-reversal-symmetry-breaking superconductor with nodes, \emph{i.e.} a Weyl superconductor.
\end{abstract}

\maketitle

%
%%
%%%
\section{Introduction}
\label{sec:Introduction}
%%%
%%
%

The widely-celebrated Bardeen Cooper Schrieffer (BCS) theory describes superconductivity deriving from Cooper pairs in a zero relative angular momentum ($s$-wave) state~\cite{BCSTheory}. It was later established that superconductivity could arise from purely repulsive electron interactions~\cite{KohnLuttinger}, in stark contrast to the phonon-mediated coupling of BCS theory. Cooper pairs originating from repulsive interactions are typically non-$s$-wave, and chiral complex combinations, \emph{e.g.}\@ $p_x \pm ip_y$, often lower the free energy~\cite{Kallin}. The list of established unconventional superconducting materials is rapidly growing. Well-known examples include the $d$-wave high $T_c$ cuprate compounds. Other extensively-studied unconventional superconductors include the perovskite Sr$_2$RuO$_4$~\cite{SCNature, StrontiumRuthenateMaeno2003, Sr2RuO4and3He}, the heavy-fermion UPt$_3$~\cite{Upt3Overview, FSUPt3}, and SrPtAs~\cite{SrPtAsdwave}, thought to have $p$-, $f$- and $d$-wave order, respectively. 

At weak coupling the repulsive Hubbard model has been used as an illustrative platform to study unconventional superconductivity in two dimensions~\cite{PairinstabChubukov, PRBRGFlowHubbard, PhaseDiagramSimkovic, EPLRPA, PRA2DHubbard, PhaseDiagramKreisel}. For the 3D simple cubic lattice the weak-coupling ground state phases have been established~\cite{dwaveinstabHirsch, PRBRGFlowHubbard}. However, with the out-of-plane hopping strength $t_{\perp}$ being different from $0$ (2D) and different from the in-plane hopping strength $t_{\parallel}$ (simple cubic lattice), much less is known. In particular, making $t_{\perp}$ finite but small makes the Fermi surface a corrugated cylinder at low filling. The effect of corrugation on the order parameter has not been explored within the weak-coupling scheme. The problem has been treated within the mean field approximation~\cite{TakimotoPRB} and has been discussed in terms of the thermal Hall conductivity~\cite{SigristChiralSCnodes}. 

In this paper we demonstrate the importance of corrugation effects in unconventional superconductors by establishing the superconducting weak-coupling phase diagram for a repulsive Hubbard model in $(\mu, t_{\perp})$ space, where $\mu$ is the chemical potential and $t_{\perp}$ the out-of-plane hopping. We consider a tight-binding single-band model with a minimal number of free parameters. Our model connects the two- to the three-dimensional case, and it spans four Fermi surface topologies below half-filling. We employ a weak-coupling procedure which allows us to calculate the order parameter from first principles~\cite{PRBRGFlowHubbard}. The method is considered exact in the limit $U/t \to 0$ assuming that $\omega \gg U^2/t$, where $U$ is the on-site interaction strength, $\omega$ the electronic bandwidth, and $t$ is the scale of the hopping terms (see Appendix \ref{app:MethodDetails} for further details). In various regions of the phase diagram we find that the gap has point or line nodes. Surprisingly, we find corrugation-induced nodes close to the cylindrical limit, challenging the typical view of the chiral $p_x + ip_y$ phase being uniform and fully gapped over the Fermi surface.

%
%%
%%%
\section{Model Hamiltonian and the Weak-Coupling Approach}
\label{sec:Model}
%%%
%%
%

We set up a nearest-neighbor tight-binding model to address the impact of 3D effects in unconventional superconductors. Electrons at chemical potential $\mu$ hop on a primitive tetragonal lattice with an out-of-plane hopping strength $t_{\perp}$ and in-plane hopping strength $t_{\parallel}$. We assume that the electrons interact via a weak repulsive on-site interaction $U > 0$,
\begin{equation}
\begin{aligned}
H &= -t_{\parallel} \sum_{\sigma, \langle i, j \rangle_{\parallel}} c_{i \sigma}^{\dagger} c_{j \sigma}^{\phantom{\dagger}} -t_{\perp} \sum_{\sigma, \langle i, j \rangle_{\perp}} c_{i \sigma}^{\dagger} c_{j \sigma}^{\phantom{\dagger}} \\
& \hspace{10pt} - \mu \sum_{\sigma, i} n_{i \sigma} + U \sum_{i} n_{i \uparrow} n_{i\downarrow},
\end{aligned}
\label{eq:Hamiltonian}
\end{equation}
where $c^{\left(\dagger\right)}_{i\sigma}$ annihilates (creates) an electron with spin $\sigma \in \lbrace \uparrow, \downarrow \rbrace$ on site $i$, $n_{i \sigma} = c^{\dagger}_{i \sigma} c_{i \sigma}$ is the number operator, and $\langle \cdot \rangle_{\parallel}$ and $\langle \cdot \rangle_{\perp}$ denote in-plane and out-of-plane nearest neighbor sites, respectively. Fourier transforming the Hamiltonian yields the single-particle dispersion
\begin{equation}
\xi_{\bo{k}} = -2 t_{\parallel} \left(  \cos{k_x} + \cos{k_y}  \right) -2 t_{\perp} \cos{k_z} - \mu.
\label{eq:Dispersion}
\end{equation}

We assume in the following that $U/t_{\parallel} \ll 1$, and the superconducting order is calculated perturbatively by treating the interaction to one-loop order, $\pazocal{O}(U^2)$ (see Appendix \ref{app:MethodDetails}) \cite{PRBRGFlowHubbard, Scaffidi2017weak}. In this framework, the order parameter is determined from the effective particle-particle vertex. In the triplet ($t$) (singlet ($s$)) channel, this reads
\begin{equation}
\int_{S_{F}} \f{\D^2 k'}{\lvert S_{F} \rvert} g^{t/s}_{\bo{k}, \bo{k}'} \psi^{(n)}_{t/s;\bo{k}'} = \lambda_n \psi^{(n)}_{t/s; \bo{k}},
\label{eq:Eigenmodeequation}
\end{equation}
where the integral is over the Fermi surface and $\lvert S_{F} \rvert$ is its area. The matrix $g^{t/s}_{\bo{k}, \bo{k}'}$ is the dimensionless two-particle vertex and is given by
\begin{equation}
g^{t/s}_{\bo{k}, \bo{k}'} = \rho_0 U^2 \sqrt{\frac{\bar{v}_F}{v_F(\bo{k})} } \Gamma^{t/s}_{\bo{k},\bo{k}'} \sqrt{\frac{\bar{v}_F}{v_F(\bo{k}')} }.
\label{eq:gmatrixtriplet}
\end{equation}
% \chi(\bo{k}-\bo{k}') 
Here, $\Gamma^{t}_{\bo{k},\bo{k}'} = -\chi(\bo{k}-\bo{k}')$ and $\Gamma^{s}_{\bo{k},\bo{k}'} = \chi(\bo{k}+\bo{k}')$ to one-loop order (Appendix \ref{app:MethodDetails}), $\bar{v}_F^{-1} = \int_{S_{F}} \frac{\D^2 k}{\lvert S_{F} \rvert} v_F(\bo{k})^{-1}$, $\rho_0 = \lim_{\lvert \bo{q} \rvert \to 0} \chi(\bo{q})$, 
\begin{equation}
\chi(\bo{q}) = -\int \frac{\D^3 p}{(2\pi)^3} \frac{f(\xi_{\bo{p}})-f(\xi_{\bo{p}+\bo{q}})}{\xi_{\bo{p}}-\xi_{\bo{p}+\bo{q}}}
\label{eq:Lindhard}
\end{equation}
is the density-density response (the Lindhard function), and $f(E)$ the Fermi-function. An eigenfunction of the integral equation corresponding to a negative eigenvalue $\lambda_n$ signals the onset of superconductivity with an order parameter 
\begin{equation}
\Delta_{t/s;\bo{k}}^{(n)} \sim T_c^{(n)} \sqrt{\f{v_F(\bo{k})}{\bar{v}_F} } \psi_{t/s;\bo{k}}^{(n)}
\label{eq:GapFunction}
\end{equation}
below the critical temperature $T_c^{(n)} \sim \omega e^{ -1/\lvert \lambda_n \rvert }$, where $\omega$ is the bare bandwidth (when $\omega \gg U^2/t$). In the numerical scheme in Sec.\@ \ref{sec:NumericalResults} the Lindhard function was regularized by adding a small imaginary contribution to the denominator of Eq.\@ \eqref{eq:Lindhard}. 

The order parameter belongs to a representation of the relevant lattice point group. In our case this is the tetragonal point group $D_{4h}$ as summarized in Table \ref{tab:Representations}. By the Pauli principle, the solution must be symmetric (antisymmetric) in the singlet (triplet) channel. We define the ground state as the order parameter with the highest $T_c$.

A useful quantity to help distinguish topological phases is the Chern number, defined for any 2D slice of the 3D Brillouin zone~\cite{TopInsandSC}. We choose to define it, via Stokes' theorem, in terms of $k_z$ slices of the Fermi sea, measuring the winding of the order parameter phase,
\begin{equation}
C(k_z) = \f{1}{2\pi}\oint_{\mathrm{FS}(k_z)} \D\bo{k} \cdot \nabla \mathrm{arg}(\Delta_{t/s;\bo{k}}).
\label{eq:ChernNumber}
\end{equation}
This quantity is a topological invariant as long as the gap is nonzero along the integrated path. As $k_z$ is smoothly varied, the Chern number can jump up or down by an integer when the integration path passes through a point node. 

\begin{table}[h!tb]
\caption{Even- ($g$) and odd- ($u$) parity representations of the tetragonal point group $D_{4h}$~\cite{PointGroupSymmetries, SigristPheno}. Here, one should associate $x$ with a function that transforms like $\sin(k_x)$ under the group operations, $x^2$ with a function that transforms like $\cos(k_x)$, and similar associations for $y$ and $z$. Basis functions in braces are degenerate.}
\begin{center}
\begin{tabular}{p{1.2cm} p{2.5cm} }
\hline
 Rep. & Basis functions \\ \hline
 $A_{1g}$ & $s$ or $d_{2z^2-x^2-y^2}$ \\
 $A_{2g}$ & $g_{xy(x^2-y^2)}$ \\
 $B_{1g}$ & $d_{x^2-y^2}$ \\
 $B_{2g}$ & $d_{xy}$ \\
 $E_{g}$ & $\big\lbrace d_{yz}, d_{zx} \big\rbrace$ \\
 $A_{1u}$ & $h_{xyz(x^2-y^2)}$ \\
 $A_{2u}$ & $p_z$ \\
 $B_{1u}$ & $f_{xyz}$ \\
 $B_{2u}$ & $f_{z(x^2-y^2)}$ \\
 $E_{u}$ & $\big\lbrace p_x, p_y \big\rbrace$ \\
  \hline
\end{tabular}
\end{center}
\label{tab:Representations}
\end{table}

%
%%
%%%
\section{Numerical results}
\label{sec:NumericalResults}
%%%
%%
%

Initial attempts at establishing the phase diagram of the two-dimensional ($t_{\perp} = 0$) repulsive Hubbard model at weak coupling~\cite{PairinstabChubukov, PhaseDiagramHlubina} were later refined~\cite{PRBRGFlowHubbard, PhaseDiagramSimkovic} and approached with the random phase approximation~\cite{EPLRPA, PRA2DHubbard, PhaseDiagramKreisel}. Similarly, the phases of the simple cubic lattice repulsive Hubbard model ($t_{\perp} = t_{\parallel}$) have also been established (still at weak coupling)~\cite{dwaveinstabHirsch, PRBRGFlowHubbard}. Remaining unexplored, however, is the transition between two and three dimensions, $0 < t_{\perp} < t_{\parallel}$, and the anisotropic cases $t_{\perp} > t_{\parallel}$.

To establish the complete phase diagram we apply the weak-coupling scheme and discretize the Fermi surface, typically using $3000$ to $4000$ points, such that Eq.\@ \eqref{eq:Eigenmodeequation} becomes a regular matrix eigenvalue problem. The susceptibility (Eq.\@ \eqref{eq:Lindhard}) is calculated at $T = 0$ with a uniform integration mesh and regularization of the Lindhard function.

\subsection{The phase diagram}

\begin{figure}[h!tb]
\centering
\includegraphics[width=\linewidth]{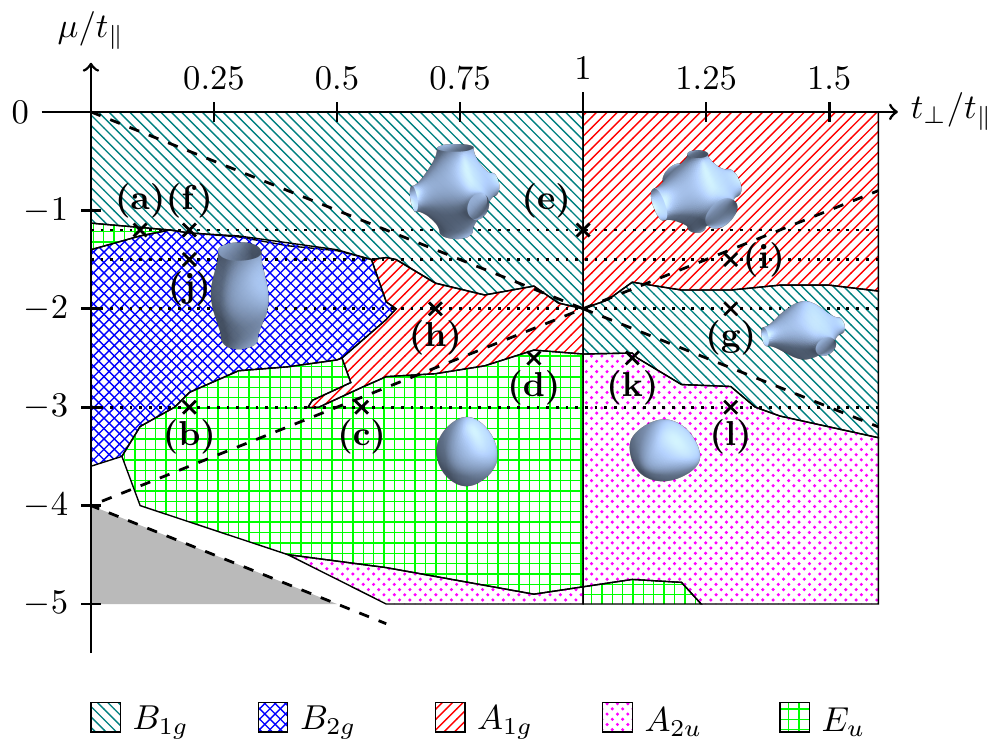} 
\caption{(Color online). Phase diagram in the superconducting phase of the repulsive single-band Hubbard model in Eq.\@ \eqref{eq:Hamiltonian}, covering four Fermi surface topologies with samples displayed in blue. In the gray shaded area there is no Fermi surface. The black dashed lines mark the van Hove singularities, the black crosses with labels refer to the order parameters examined in Fig.\@ \ref{fig:OPexamples}, and the horizontal dotted lines show the cuts displayed in Fig.\@ \ref{fig:CouplingStrengths}. In the white regions the Fermi surface is too small for the numerical scheme to be trusted.}
\label{fig:PhaseDiagram}
\end{figure}

\begin{figure}[h!tb]
	\centering
	\subfigure[]{\includegraphics[width=0.84\linewidth]{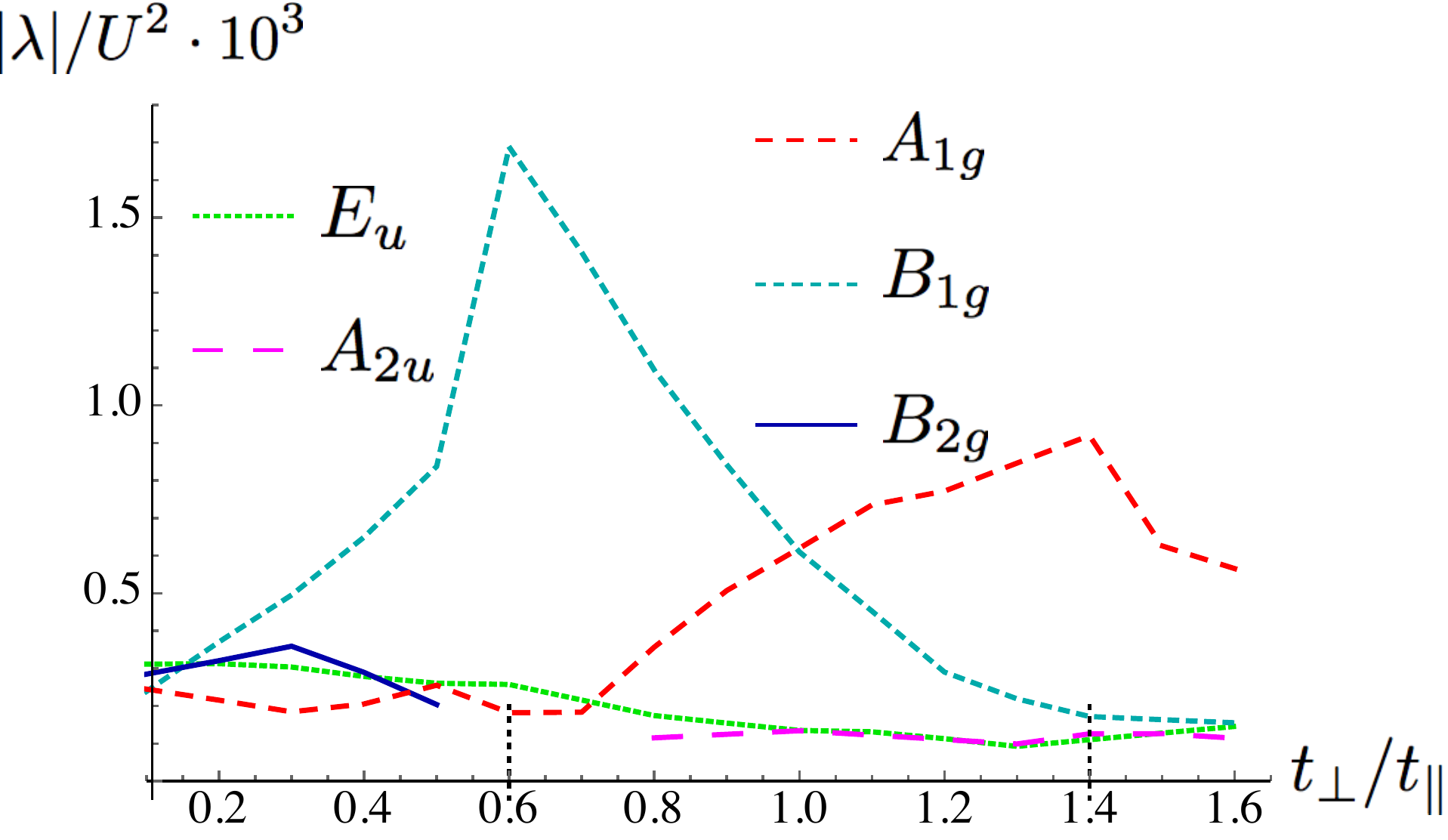} }\quad \subfigure[]{\includegraphics[width=0.84\linewidth]{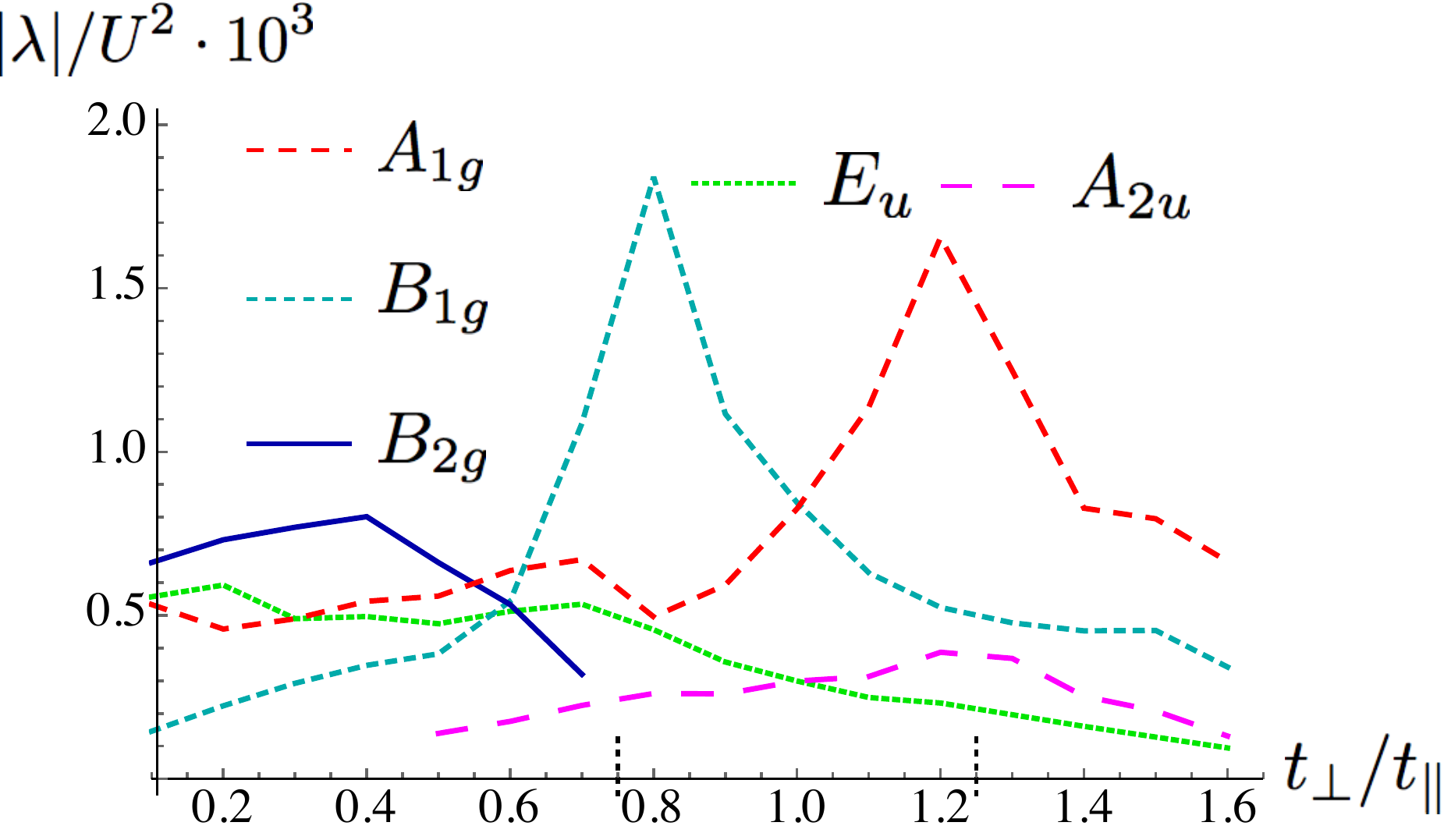}}\quad \subfigure[]{\includegraphics[width=0.86\linewidth]{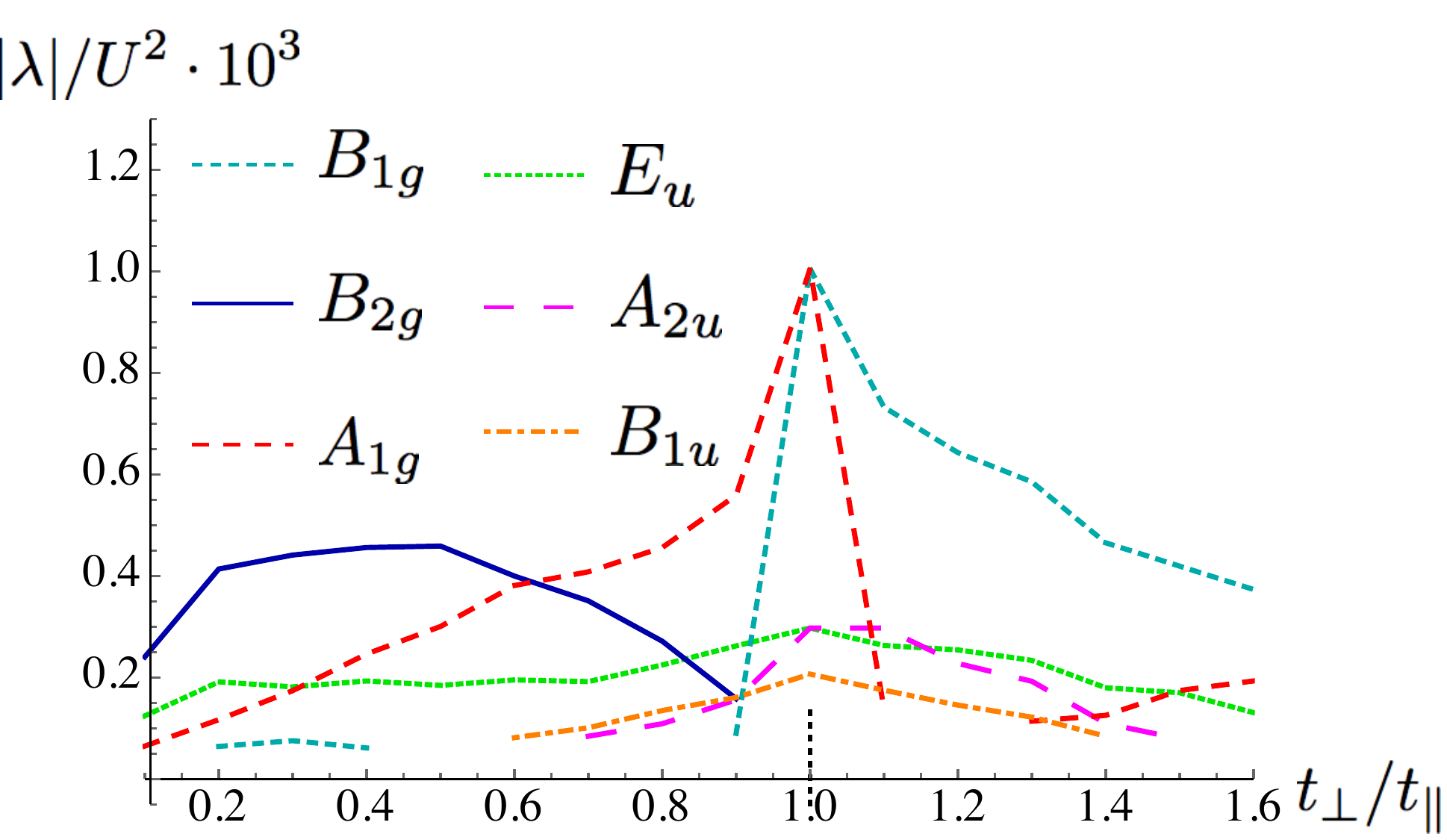}}
\quad \subfigure[]{\includegraphics[width=0.86\linewidth]{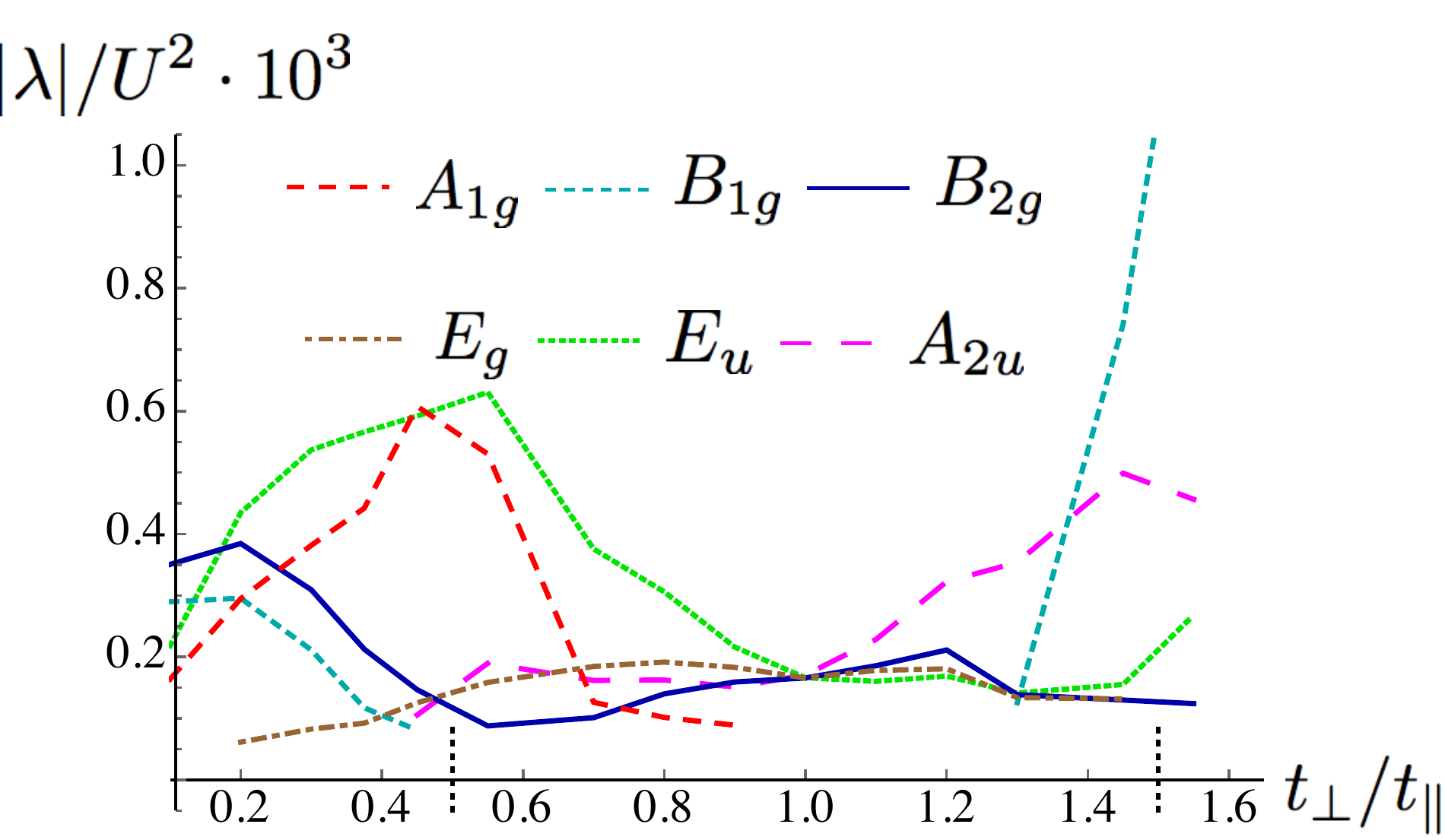}}
\caption{(Color online). Competing coupling strengths as a function of $t_{\perp}$ for $\mu = -1.2, -1.5, -2.0, -3.0$ in (a), (b), (c), (d), respectively. Only the most relevant representations are shown. The vertical dotted lines indicate the van Hove points.}
	\label{fig:CouplingStrengths}
\end{figure}

In Fig.~\ref{fig:PhaseDiagram} we show the phase diagram obtained by varying the chemical potential $\mu$ and out-of-plane hopping $t_\perp$. The diagram was constructed by identifying the most negative eigenvalue of $g_{\bo{k},\bo{k}'}$, in both the triplet and the singlet sector, with a resolution of $\delta t_{\perp} = 0.1$ for a range of $\mu$ with $t_{\parallel} = 1.0$ fixed throughout. The corresponding coupling strengths to second order in the Hubbard interaction for the closest competing point group representations are shown for a selection of chemical potentials in Fig.\@ \ref{fig:CouplingStrengths}. 

\begin{figure*}[h!bt]
	\centering
\subfigure[]{\includegraphics[width=0.23\linewidth]{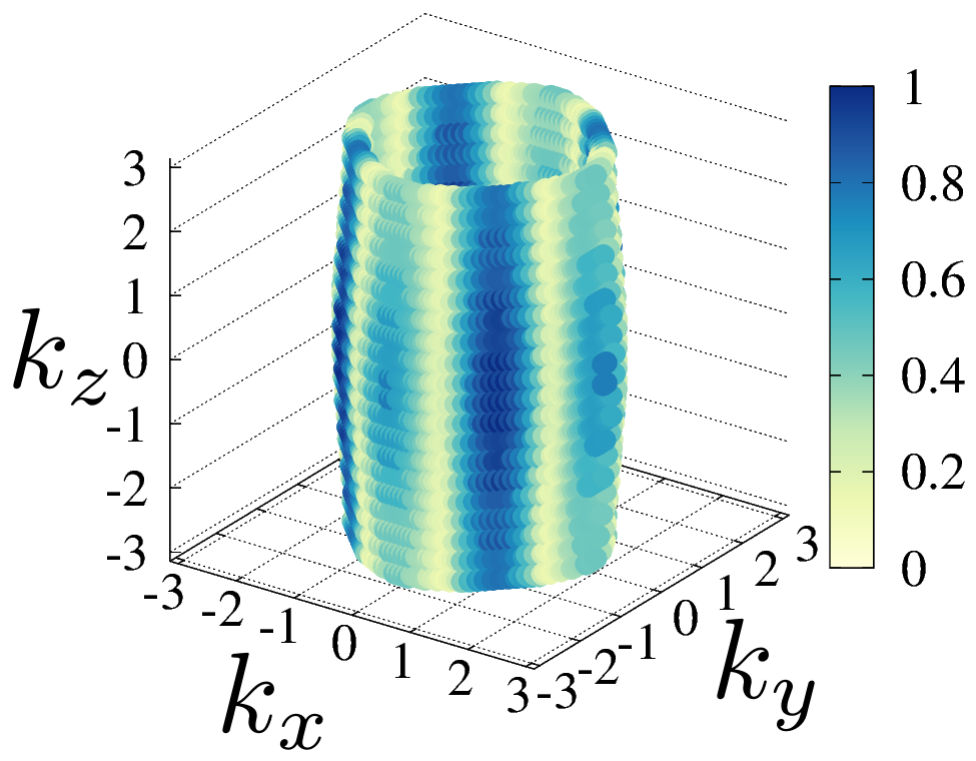}}\quad \subfigure[]{\includegraphics[width=0.23\linewidth]{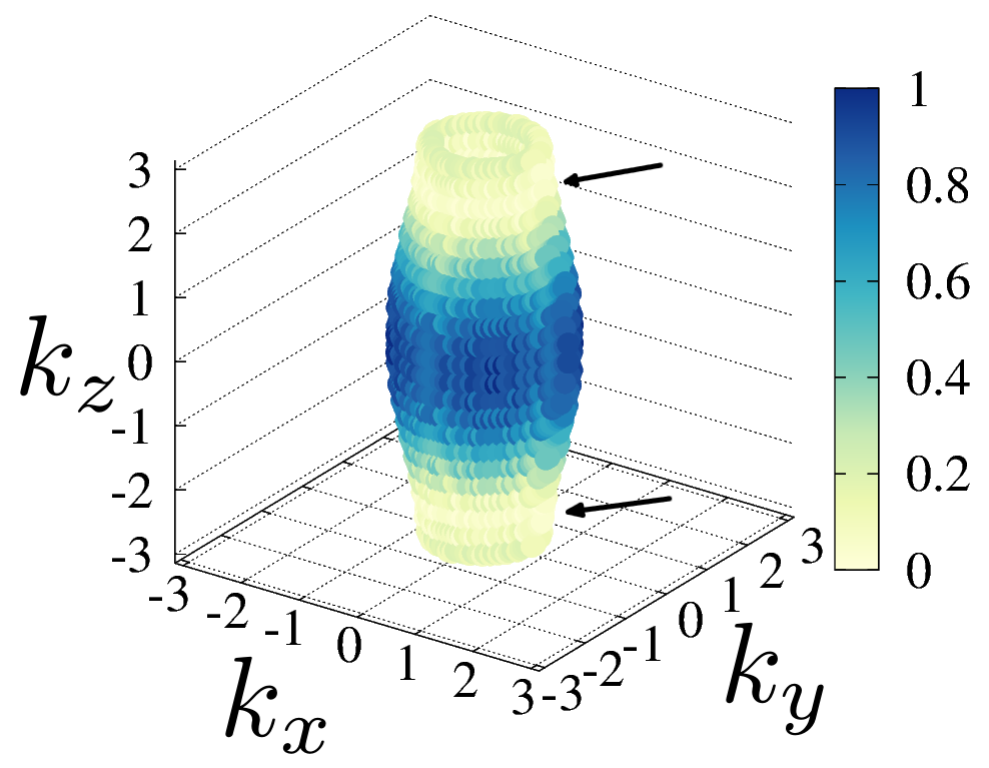}} \quad \subfigure[]{\includegraphics[width=0.23\linewidth]{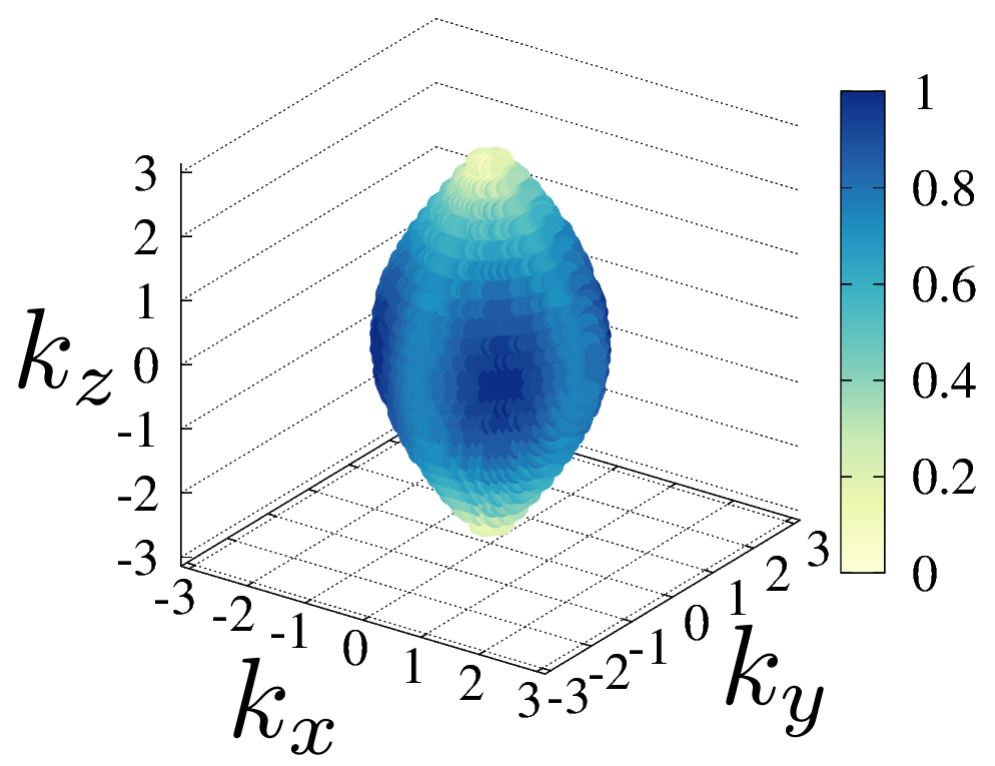}} \quad \subfigure[]{\includegraphics[width=0.23\linewidth]{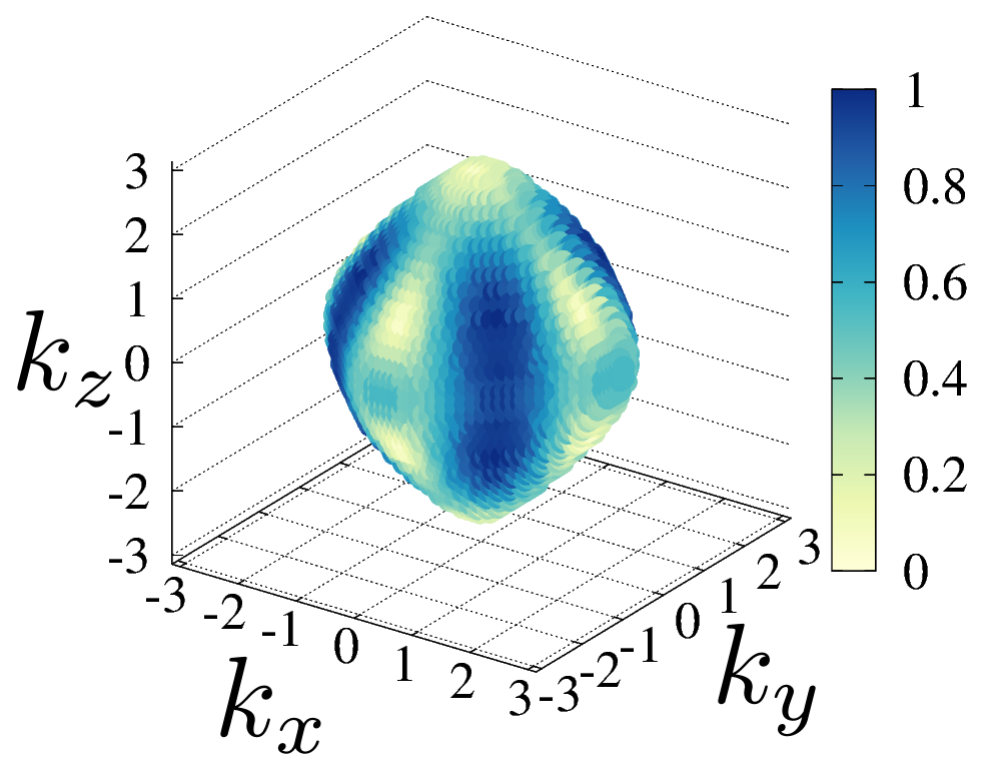}} \quad \subfigure[]{\includegraphics[width=0.23\linewidth]{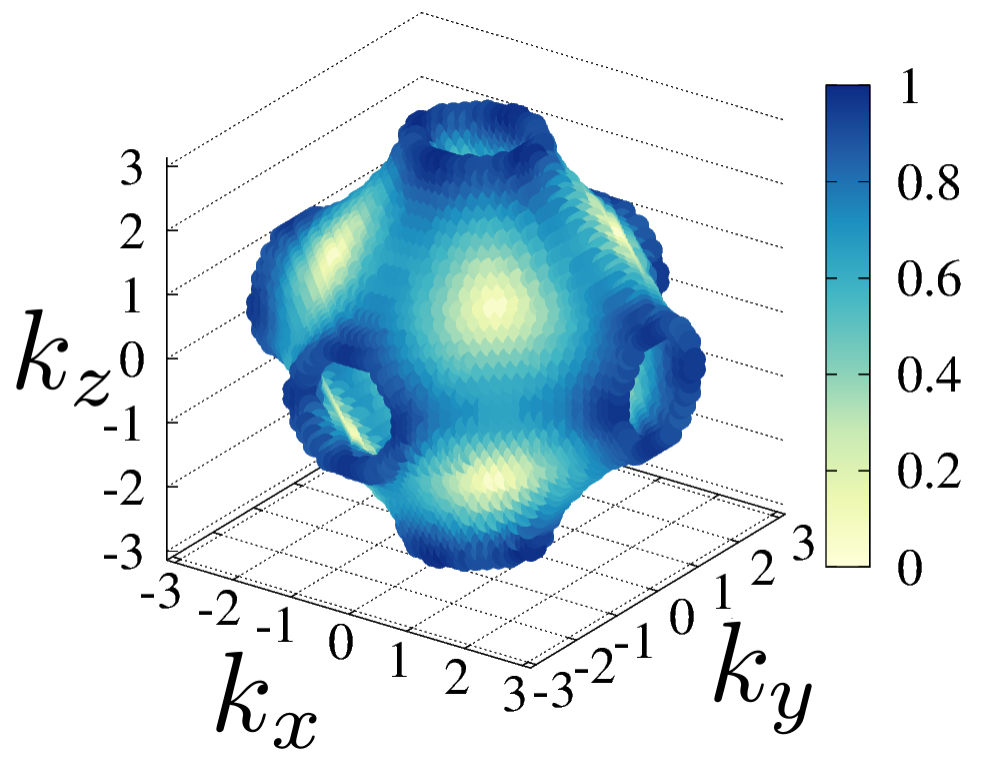}} \quad \subfigure[]{\includegraphics[width=0.23\linewidth]{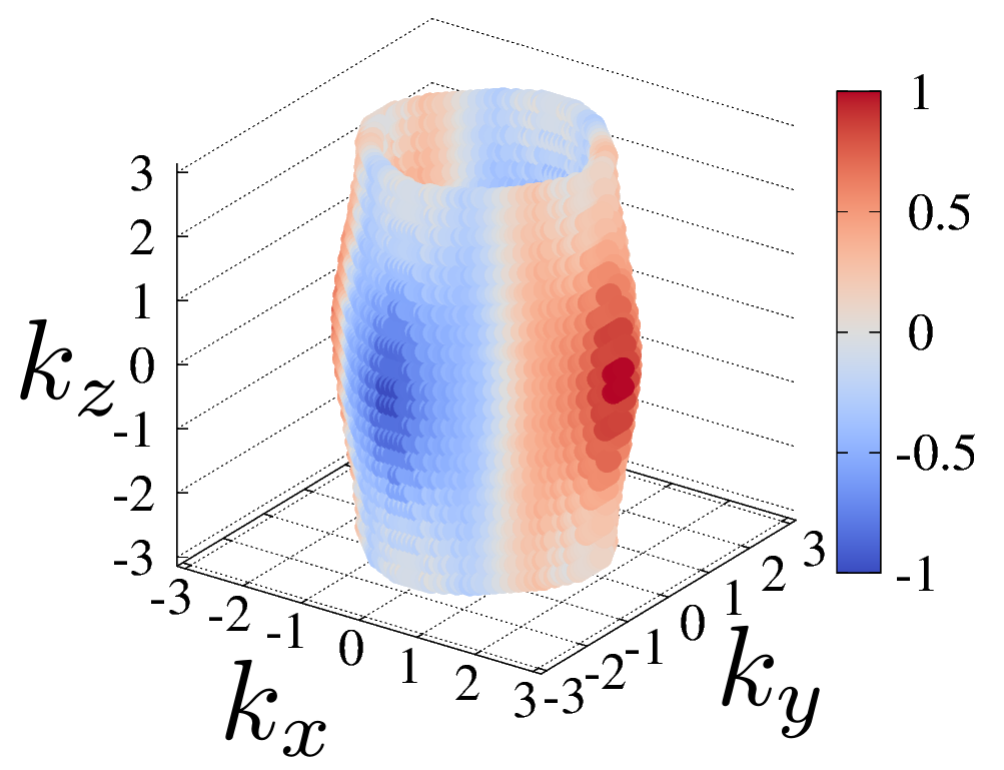}} \quad \subfigure[]{\includegraphics[width=0.23\linewidth]{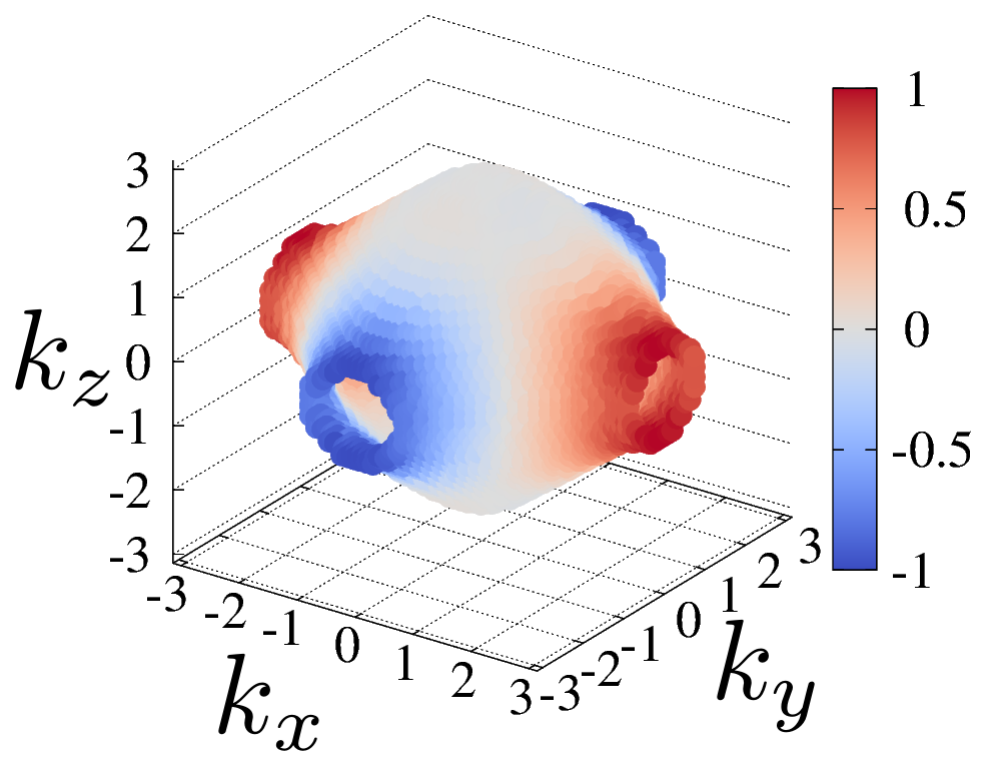}} \quad \subfigure[]{\includegraphics[width=0.23\linewidth]{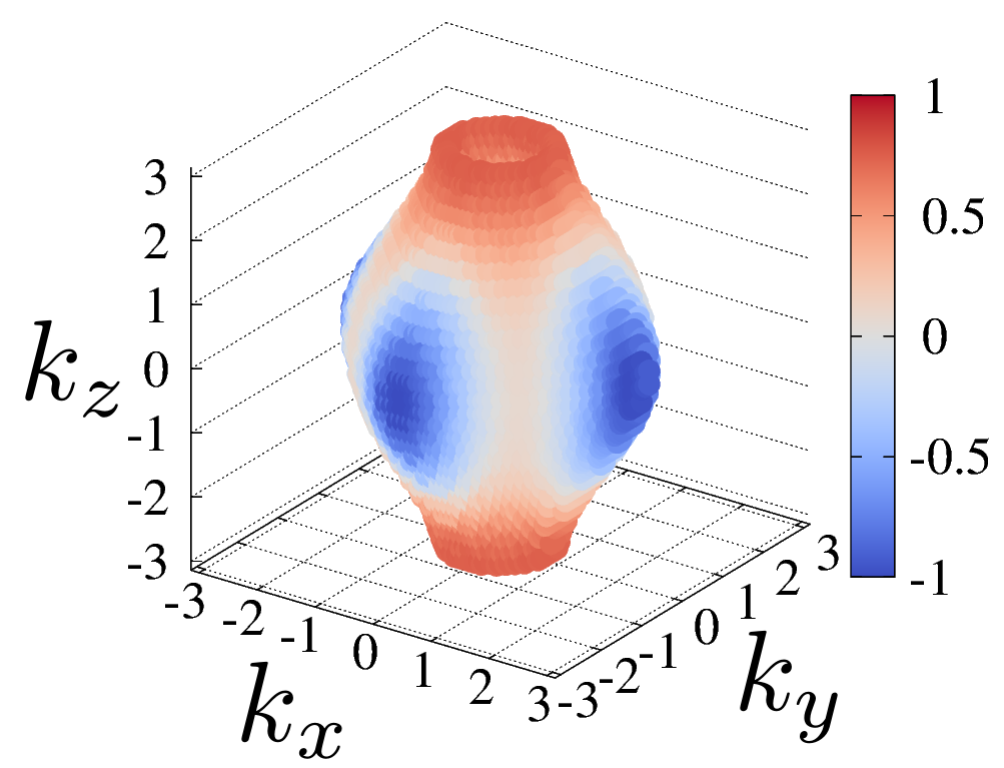}} \quad \subfigure[]{\includegraphics[width=0.23\linewidth]{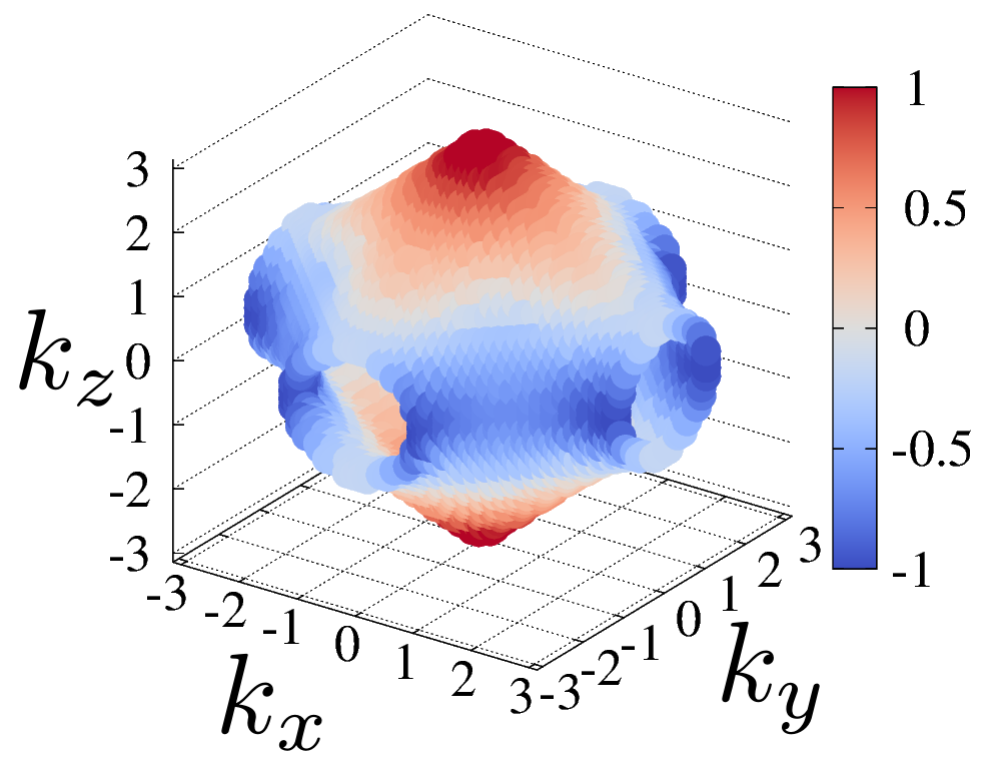}} \quad \subfigure[]{\includegraphics[width=0.23\linewidth]{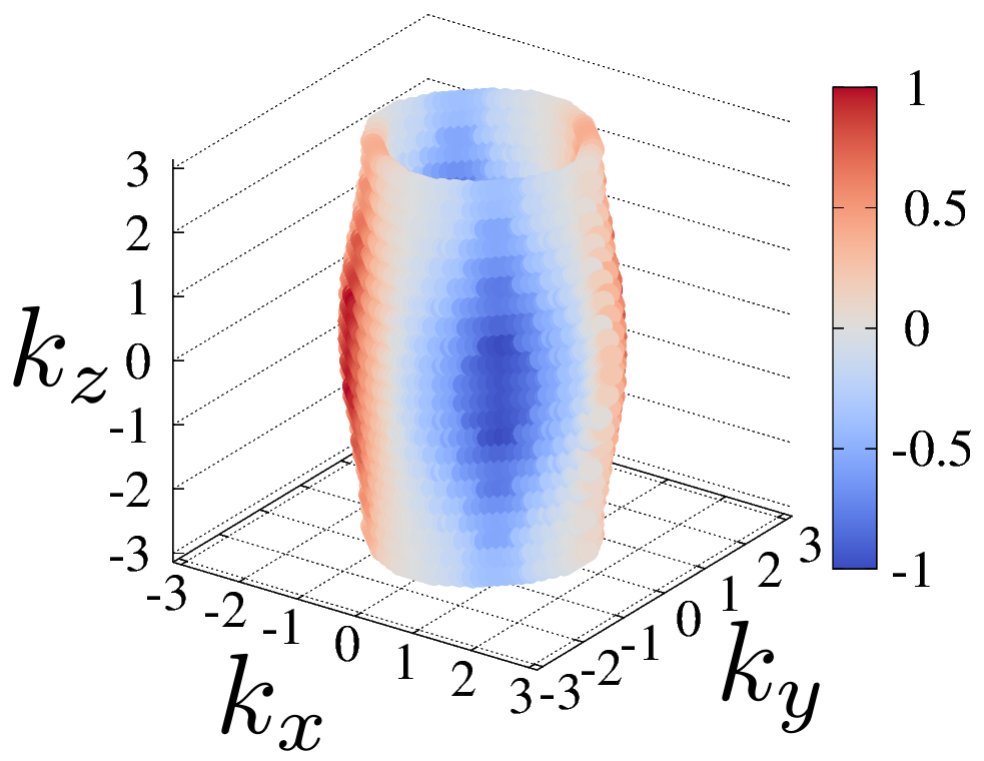}} \quad \subfigure[]{\includegraphics[width=0.23\linewidth]{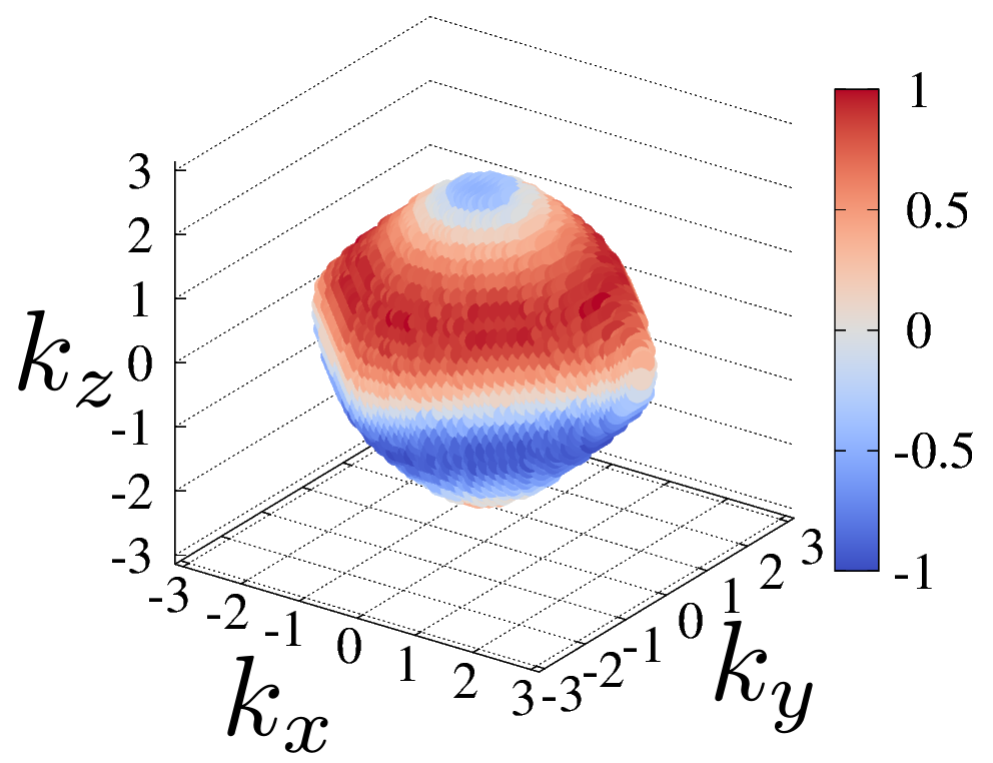}} \quad \subfigure[]{\includegraphics[width=0.23\linewidth]{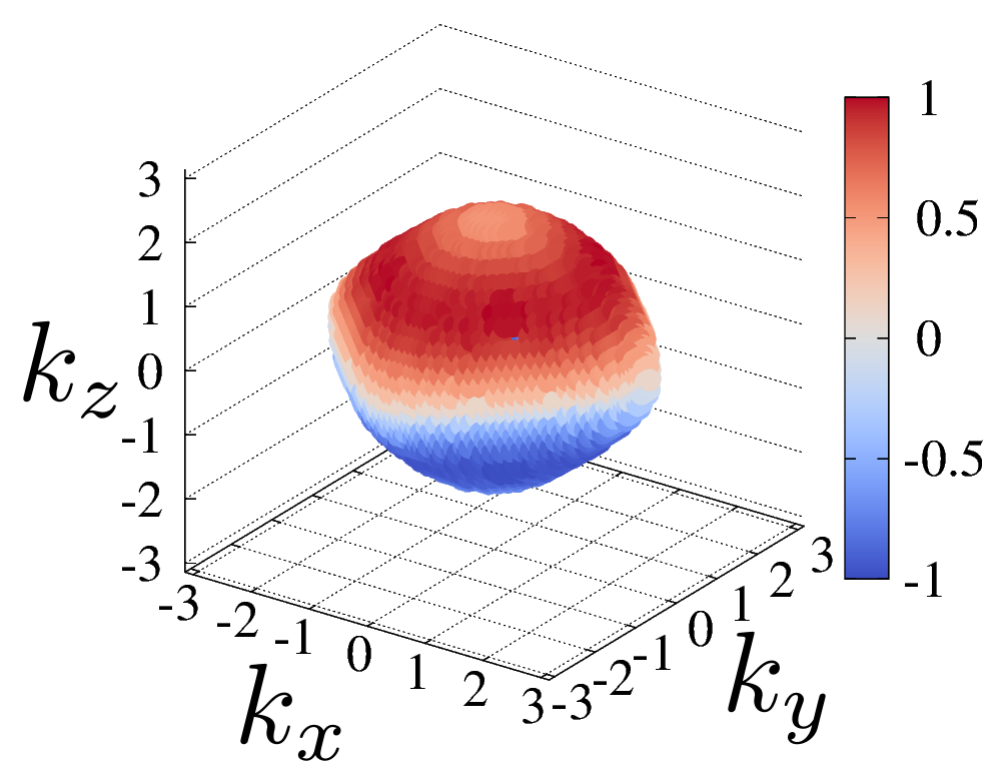}} 
\caption{(Color online). (a) -- (e) Magnitudes of some time-reversal-breaking order parameters found in this study. (f) -- (l) Examples of non-chiral order parameters also found. The respective positions in the phase diagram are shown in Fig.\@ \ref{fig:PhaseDiagram}. (a) Fully gapped $p_x + ip_y$ dominated by the lattice harmonics $\sin(3k_x) + i\sin(3k_y)$ ($C = -3$). (b) $p_x+ip_y$ with two horizontal line nodes (arrows) and $C = 1$ on both sides of the line nodes. (c) $p_x+ip_y$ with two point nodes and $C = 1$. (d) $p_x+ip_y$ with a total of ten point nodes. Here, $C = -3$ is realized for small $\lvert k_z \rvert$ and $C = +1$ is realized for larger $\lvert k_z \rvert$. (e) $d_{x^2-y^2} + i d_{2z^2-x^2-y^2}$ with eight point nodes and $C = 0$ throughout. (f), (g) $d_{x^2-y^2}$ for two different Fermi surface topologies. (h), (i) $d_{2z^2-x^2-y^2}$ for two different Fermi surface topologies. (j) $d_{xy}$ with a weak $k_z$ dependency. (k) $p_z$ with two accidental horizontal line nodes at finite $k_z$. (l) $p_z$ with dominant fundamental lattice harmonic.}
\label{fig:OPexamples}
\end{figure*}

Order parameters belonging to five different representations of  the point group $D_{4h}$ are seen to be realized as the highest $T_c$ order within the parameter window $t_{\perp} \in [0.0, 1.6]$ for $\mu \in [-5, 0]$. We should emphasize that the gap is always given by a linear combination of lattice harmonics that lie in the same given irreducible representation. While in certain cases, this linear combination is dominated by the fundamental lattice harmonic (for example $\cos(k_x) - \cos(k_y)$ for $B_{1g}$), this is in general not the case, and the basis functions given in Table I should therefore not be understood as accurate descriptions of the gap. Indeed, even within the same representation we find a large variety in the detailed structure of the gap, for example in terms of accidental nodes, as illustrated in Fig.\@ \ref{fig:OPexamples}. 

Comparing with previously established results, we recover the transition from $\lbrace d_{x^2-y^2}, d_{2z^2-x^2-y^2} \rbrace$ to $\lbrace p_x, p_y, p_z\rbrace$ at $\mu = -2.46$ ($n = 0.32$) for $t_{\perp} = t_{\parallel}$~\cite{PRBRGFlowHubbard}. At $\mu \sim -3.0$ we note that the $E_g$ and $B_{2g}$ states, which were left out in Ref.\@ \onlinecite{PRBRGFlowHubbard}, are practically degenerate with the $E_u$ states around $t_{\perp} = t_{\parallel}$ (Fig.\@ \ref{fig:CouplingStrengths} (d)). Expectedly, the order parameters belonging to the same irreducible representation of $O_h$, the octahedral point group, become degenerate in the case of $t_{\perp} = t_{\parallel}$. Along the line $t_{\perp} = 0$ we also recover the expected phases of the two-dimensional model~\cite{PhaseDiagramSimkovic}: As the chemical potential is lowered, a transition from $B_{1g}$ to $E_u$ occurs at $\mu = -1.15$ ($n = 0.58$), but the odd-parity phase is overtaken by $B_{2g}$ already at $\mu = -1.40$ ($n = 0.51$). The $B_{2g}$ phase is well described, throughout its region in the phase diagram, by the $d_{xy}$ order parameter (see Fig.\@ \ref{fig:OPexamples} (j)).

Across the van Hove singularity at $\mu^{(x,y)} = -2t_{\perp}$ the Fermi surface changes topology due to the $k_x, k_y = \pm \pi$ zone boundaries. An enhanced density of states close to this line tends to favor gap symmetries for which the magnitude is large at the points where the Fermi surface touches the $k_x$ and $k_y$ zone boundaries, respectively. This is most notable for $d_{x^2-y^2}$ (see Fig.\@ \ref{fig:CouplingStrengths}). On the contrary, the van Hove line $\mu^{(z)} = -4t_{\parallel} + 2t_{\perp}$ tends to favor $d_{2z^2-x^2-y^2}$ which has peaks in the magnitude at the $k_z$ zone boundary. We note, however, that the $A_{1g}$ state close to the two-dimensional limit is well described by the extended $s$-wave order parameter with eight line nodes, as in Ref.\@ \onlinecite{PhaseDiagramSimkovic}.

A small pocket of $E_u$ order, dominated by the lattice harmonics $\sin(3k_x)$ and $\sin(3k_y)$ (see also Sec.\@ \ref{sec:TRSbreaking}) appears with $k_F \sim 2\pi /3 $ close to the two-dimensional limit in the phase diagram. As $t_{\perp}$ increases from zero the order parameter of this phase develops an interesting but disfavorable $k_z$ dependency, shown in Appendix \ref{app:EuFate}. As the van Hove line is approached the $E_u$ phase is quickly overtaken by the $d_{x^2-y^2}$ ($B_{1g}$) phase, see Fig.\@ \ref{fig:CouplingStrengths} (a).

\begin{figure}[h!tb]
	\centering
\subfigure[]{\includegraphics[width=0.47\linewidth]{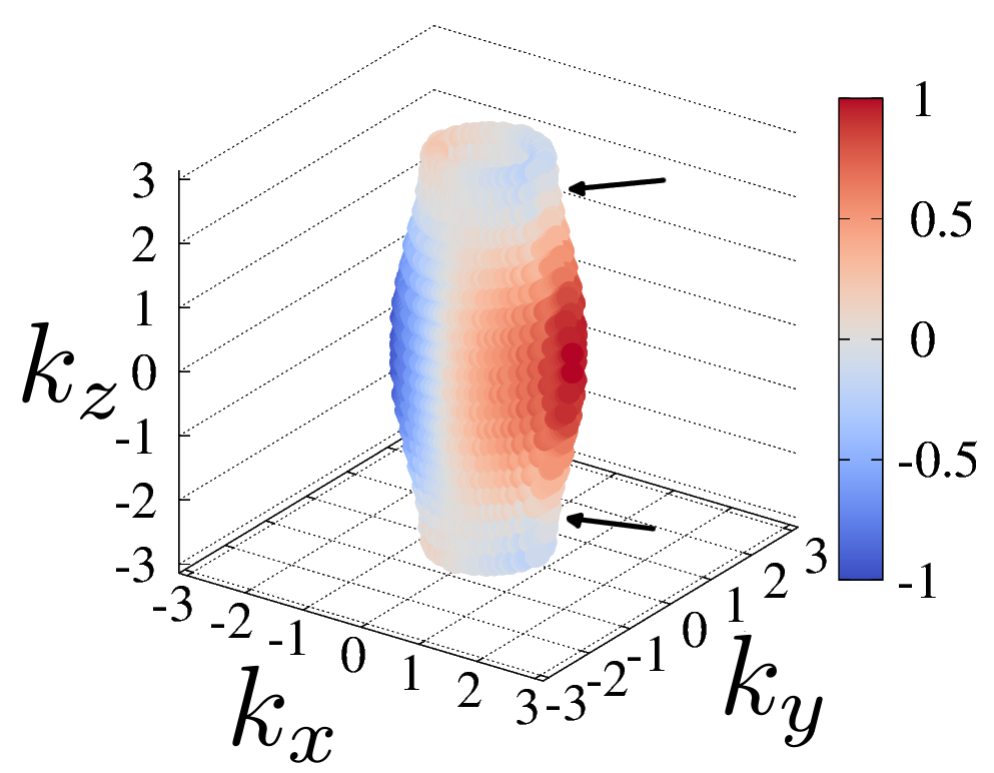}}\quad \subfigure[]{\includegraphics[width=0.47\linewidth]{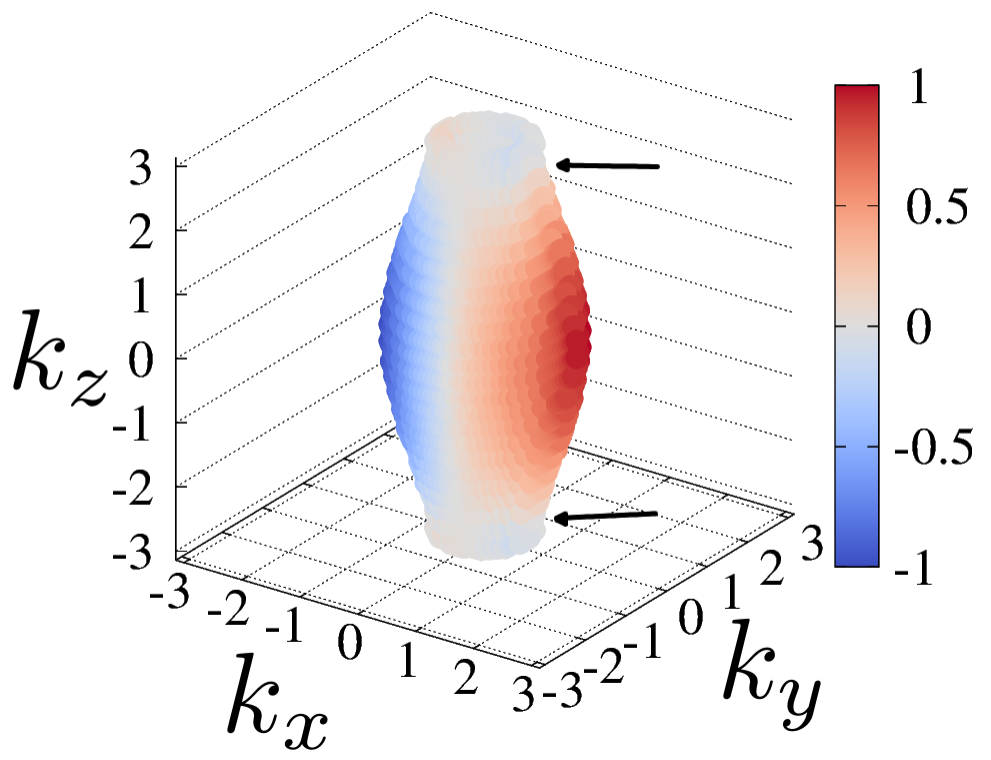}} \quad \subfigure[]{\includegraphics[width=0.47\linewidth]{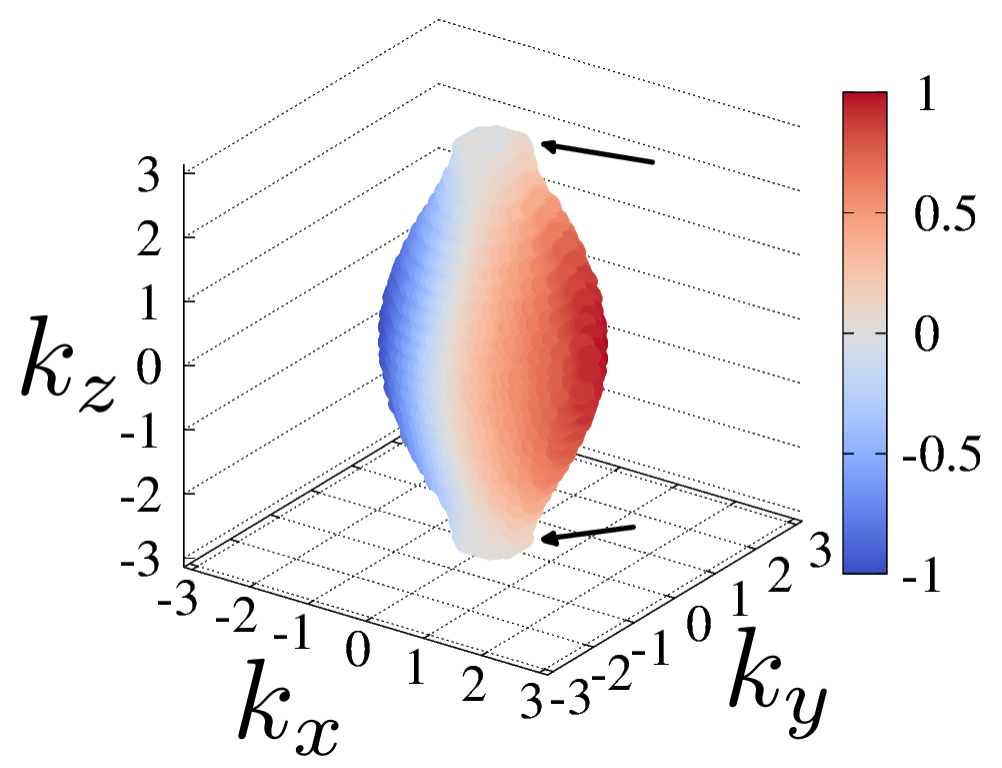}} \quad \subfigure[]{\includegraphics[width=0.47\linewidth]{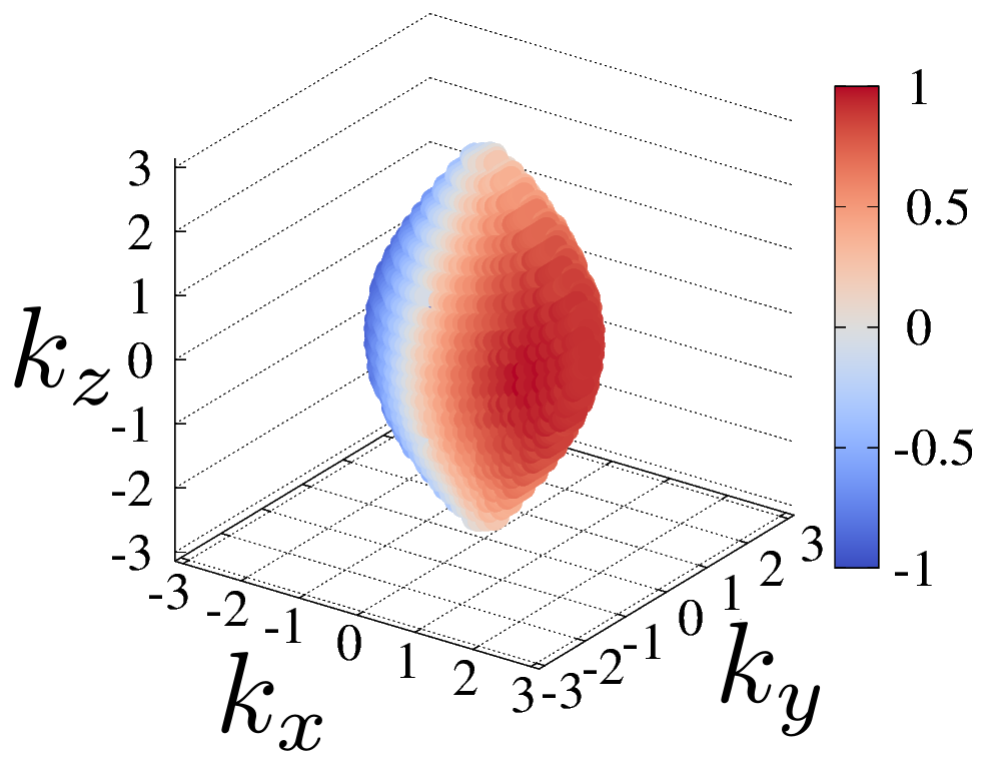}}
\caption{(Color online). The $p_x$ ($E_u$) order parameter at $\mu = -3$ for $t_{\perp}$ being (a) $0.2$, (b) $0.3$, (c) $0.45$, and (d) $0.55$. The arrows indicate the placement of the horizontal line nodes.}
\label{fig:OPpxexampleslowfilling}
\end{figure}

At lower filling an $E_u$ phase of rich structure as a function of $t_{\perp}$ emerges, see Fig.\@ \ref{fig:OPexamples} (b) -- (d). Focusing on the $\mu = -3.0$ line for concreteness (Fig.\@ \ref{fig:CouplingStrengths} (d)), the $p$-wave order parameter realized for $ 0.17 < t_{\perp} \leq 1.0$ changes its nodal nature at two points. For $t_{\perp} < 0.5$ the gap has two horizontal line nodes at some $\pm k_z'$, where $k_z'$ approaches $\pi$ as $t_{\perp}$ approaches the van Hove point. This is shown for the $p_x$ component in Fig.\@ \ref{fig:OPpxexampleslowfilling} (a) -- (c). In the range $0.5 < t_{\perp} \lesssim 0.8$ the (chiral) gap has two point nodes (see Fig.\@ \ref{fig:OPexamples} (c)). At $t_{\perp} \gtrsim 0.8$ the $p_x$ and $p_y$ gaps develop nodes on the points where the respective in-plane co-ordinate axes meet the Fermi surface. Further increasing $t_{\perp}$ makes these nodes grow into circle-like line nodes: making a local $k_z$ slice on the Fermi surface have a greater phase winding for the chiral combination, as displayed in Fig.\@ \ref{fig:OPexamples} (d). A similar feature is seen in the $p_z$ order favored at $t_{\perp} > 1$ at the same chemical potential (see Fig.\@ \ref{fig:OPexamples} (k) and (l)).

\subsection{Time-reversal-breaking combinations}
\label{sec:TRSbreaking}

In regions where two or more orders are degenerate, complex combinations are spontaneously favored to increase the condensation energy of the superconducting state. Examples of such states are shown in Fig.\@ \ref{fig:OPexamples} (a) -- (e).

The combination $d_{x^2-y^2} + id_{2z^2-x^2-y^2}$, as displayed in Fig.\@ \ref{fig:OPexamples} (e), has eight robust point nodes at $\lvert k_x \rvert = \lvert k_y \rvert = \lvert k_z \rvert$. For all gapped $k_z$ slices the gap has Chern number $0$. This complex combination (or $d - id$) is favored along all of $t_{\perp} = t_{\parallel}$ for $-2.46 \lesssim \mu < 0$.

A rather different, but no less exotic, phase we find is the $p$-wave phase at low filling for $t_{\perp}$ barely smaller than $t_{\parallel}$. This was briefly discussed at the end of the last subsection. The circular-like nodes emerging at small $k_z$ in the $E_u$ constituents cause the chiral combination $p_x + ip_y$ to have a total of ten point nodes and to realize both Chern numbers $C = +1$ for $\lvert k_z \rvert > k_z'$ and $C = -3$ for $\lvert k_z \rvert < k_z'$, where $k_z'$ is the smallest vertical value of the nodes, see Fig.\@ \ref{fig:OPexamples} (d). Phases such as this, with multiple Chern numbers within the same Fermi surface, are not usually appreciated in the literature where chiral $p$-wave order is typically discussed as a fully-gapped $\lvert C \rvert = 1$ phase.

At sufficiently low filling, found below around $\mu = -4.8$, the splitting of $p$-wave orders away from $t_{\perp} = t_{\parallel}$ flips such that $\lbrace p_x, p_y \rbrace$ are favored for the prolate spheroid Fermi surface, whereas $p_z$ is favored in the oblate regime. This agrees well with the low-filling limit as considered in the next section.

%
%%
%%%
\section{The low-filling limit: spheroidal Fermi surface}
\label{app:AnalyticLowFilling}
%%%
%%
%
We consider the model \eqref{eq:Hamiltonian} at sufficiently low fillings, such that the Fermi surface becomes rotationally symmetric around the $k_z$ axis. From Eq.~\eqref{eq:Dispersion} we find that it is given by $ k_x^2 + k_y^2 + k_z^2/\alpha^2 = k_F^2$, where $\alpha \equiv \sqrt{t_{\parallel}/t_{\perp}}$ and $k_F = \sqrt{4 + (\mu+2t_{\perp})/t_{\parallel}}$. For $\alpha>1$ this is a prolate spheroid, whereas for $\alpha<1$ the Fermi surface is an oblate spheroid.

\subsection{Spherical Fermi surface}
\label{sec:Spherical}
With $\alpha = 1$ in the low-filling limit the Fermi surface is spherical, yielding the well-known isotropic susceptibility $\chi(\bo{q}) = \rho_0 g(q/2k_F)$, where $g(x) = 1/2 + (1 - x^2)/(4x) \log\lvert(1+x)/(1-x)\rvert$ and $\rho_0 = k_F/(4\pi^2t_{\parallel})$. The effective (odd-parity) integral equation, $-\frac{\rho_0 U^2}{4\pi} \int \D\Omega_{\bo{k}'}  \hspace{1mm} \chi(\bo{k}-\bo{k}') \psi_{t,\bo{k}'} = \lambda \psi_{t,\bo{k}}$, has as solutions the spherical harmonics $Y_{\ell}^{m}$ for odd $\ell$ ($2\ell + 1$ degeneracy in $m$). By direct integration in the appropriate-parity sector, we obtain the first few solutions in $\ell$ shown in Table \ref{tab:SphericalEigenvalues} (\emph{cf.}\@ Ref.\@ \onlinecite{Layzer1968}). 
\begin{table}[h!tb]
\caption{Exact eigenvalues to one-loop order in the interaction with a spherical Fermi surface. The first three values here were first found in Ref.\@ \onlinecite{Layzer1968}.}
\begin{center}
\begin{tabular}{p{2.0cm} p{4.5cm}}
\hline
 Pairing & $\lambda / (\rho_0 U)^2$ \vspace{2pt} \\ \hline
 $s$-wave & $\f{1}{3}(1+ \log 4 ) \approx +0.80$  \\
 $p$-wave & $ -\f{1}{5}(\log 4 - 1 ) \approx -0.077$ \\
 $d$-wave & $ -\f{1}{105}(16 - 11 \log 4 ) \approx -0.0072$ \\
  $f$-wave & $ -\f{1}{945}( 69 \log 4-94 ) \approx -0.0018$ \vspace{3pt} \\
  \hline
\end{tabular}
\end{center}
\label{tab:SphericalEigenvalues}
\end{table}
Thus, the ground state order parameter is $p$-wave to one-loop order. The basis states $\lbrace p_x, p_y, p_z \rbrace$, \emph{i.e.}\@ the $T_{1u}$ representation of the octahedral point group $O_h$, have the same critical temperature when the Fermi surface is spherical.

\subsection{Prolate elongation}
\label{sec:Prolate}
Consider next $\alpha > 1$ at low filling, such that the Fermi surface is a prolate spheroid. We define the eccentricity of this spheroid as $\nu \equiv \sqrt{1-1/\alpha^2} \in (0,1)$. The prolate Fermi surface spheroid has an area of $\lvert S_F \rvert = 2\pi k_F^2 (1+\nu^{-1} \arcsin{\nu})$. Applying spheroidal coordinates, \emph{i.e.}\@ rescaling the axes, $\bo{p} = k_F(\sin{\theta}\cos{\phi}, \sin{\theta}\sin{\phi}, \alpha \cos{\theta})^T$, the Fermi velocity is $v_F(\bo{k}) = 2k_F t_{\parallel} \sqrt{1-\nu^2 \cos^2{\theta}}$. Combining this in Eq.\@ \eqref{eq:gmatrixtriplet} we find the pairing matrix in the prolate regime:
\begin{equation}
\begin{aligned}
g_{\bo{k},\bo{k}'}^{t, \mathrm{prolate}} &= - \rho_0 U^2 \frac{1+\nu^{-1}\arcsin{\nu}}{2(1-\nu^2)} \\
& \hspace{10pt} \times \frac{ \chi(\bo{k} - \bo{k}')}{\left[ \left(1-\nu^2 \cos^2{\theta}  \right) \left(1-\nu^2 \cos^2{\theta'} \right) \right]^{1/4}},
\end{aligned}
\label{eq:gmatrixPro}
\end{equation}
% \frac{2(1-\frac{\nu^2}{3})}{(1-\nu^2)(1+\nu^{-1}\arcsin{\nu})}
where $\chi$ has the same form as in the spherical case. Here, we made use of the expansion $\xi_{\bo{p}}-\xi_{\bo{p}+\bo{q}} = -t_{\parallel}\left[2\bo{p}\cdot \bo{q} + q^2 + \pazocal{O}(p,q)^4 \right]$, which is valid at low filling.
\begin{figure}[h!tb]
	\centering
	\subfigure[]{\includegraphics[width=0.67\linewidth]{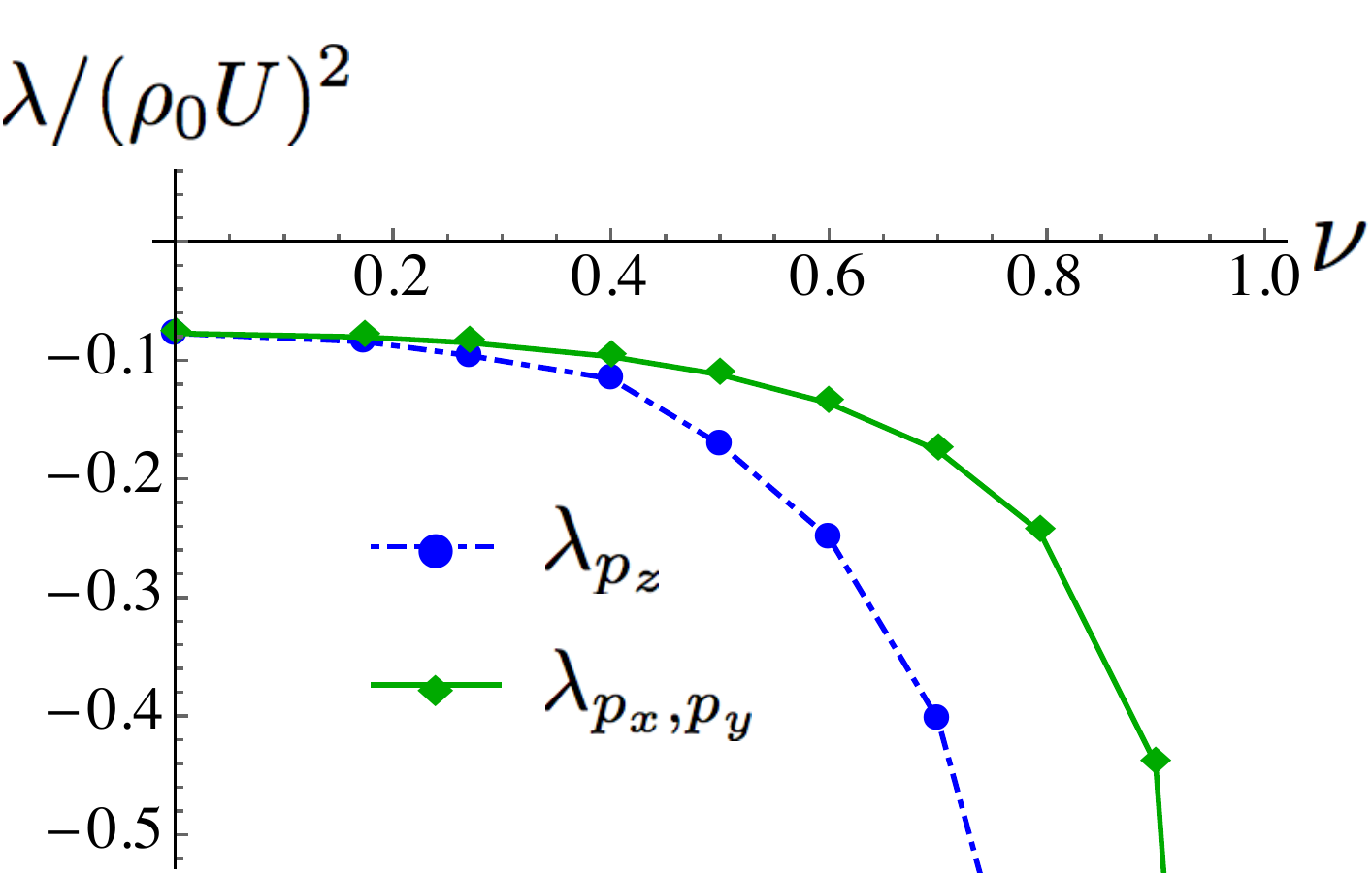} } \quad 
	\subfigure[]{\includegraphics[width=0.68\linewidth]{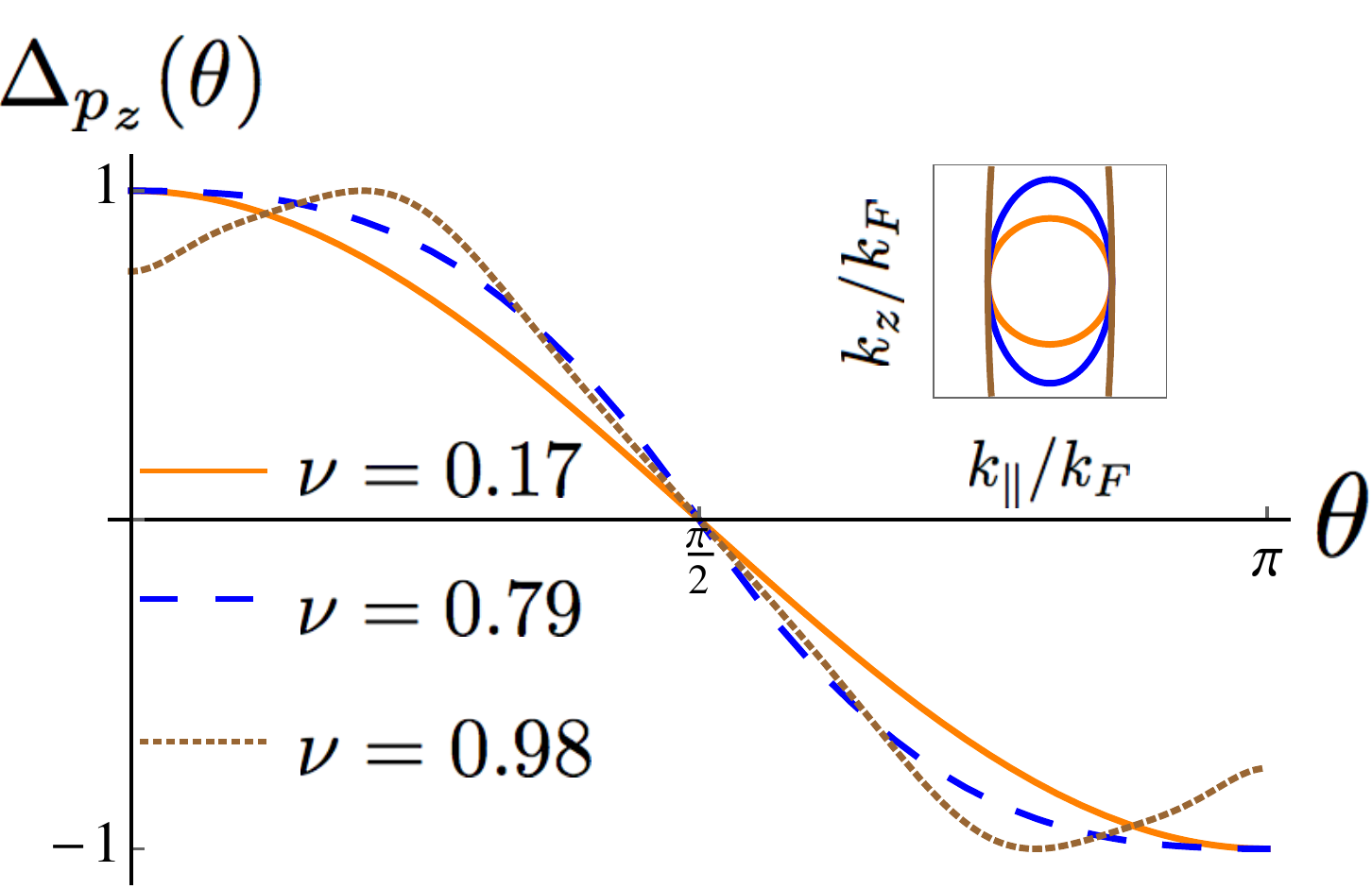} }
\caption{(Color online). (a) Pairing eigenvalues in the prolate regime by expansion in orthogonal polynomials (see Appendix \ref{app:polynomial_expansion}). (b) The ground state order parameter for three eccentricities with corresponding Fermi surfaces in the inset.}
\label{fig:ProlateNumericalSolutions}
\end{figure}
We solve Eq.\@ \eqref{eq:Eigenmodeequation} by expanding the integration kernel and its solution in orthogonal polynomials and mapping the problem to a matrix eigenvalue problem (see Appendix \ref{app:polynomial_expansion})~\cite{MorseandFeshbach}. The results are displayed in Fig.\@ \ref{fig:ProlateNumericalSolutions}, showing that the order parameter $\Delta_{p_z}$ has the highest $T_c$ for a prolate spheroidal Fermi surface. The $p$-wave value in Table \ref{tab:SphericalEigenvalues} is expectedly reassembled at $\nu = 0$. As it turns out from the numerical results in Sec.\@ \ref{sec:NumericalResults}, this splitting of $p$-wave orders applies to electron fillings around $n \lesssim 0.05$. 

\subsection{Oblate elongation}
\label{sec:Oblate}
With $\alpha < 1$ we redefine the eccentricity as $\nu \equiv \sqrt{1-\alpha^2}\in (0,1)$. Expressed in terms of $\nu$ the Fermi surface area is now $\lvert S_{F} \rvert = 2\pi k_F^2( 1 + \frac{1}{2\nu} \log{\frac{1+\nu}{1-\nu}} )$, and the Fermi velocity is $v_F(\bo{k}) = 2k_F t_{\parallel} \sqrt{1+\nu^2/(1-\nu^2) \cos^2{\theta}}$ in rescaled coordinates. This leads to the pairing matrix in the oblate regime:
\begin{equation}
\begin{aligned}
g_{\bo{k},\bo{k}'}^{t, \mathrm{oblate}} &= - \rho_0 U^2 \f{1-\nu^2}{2} \big( 1 + \frac{1}{2\nu} \log{\frac{1+\nu}{1-\nu} } \big)  \\
& \hspace{-20pt} \times \frac{ \chi(\bo{k}-\bo{k}') }{\left[ \left( 1 + \f{\nu^2}{1-\nu^2} \cos^2{\theta} \right)  \left( 1 + \f{\nu^2}{1-\nu^2} \cos^2{\theta'}  \right) \right]^{1/4}}.
\end{aligned}
\label{eq:gmatrixOb}
\end{equation}
% \frac{2(1-\frac{2}{3}\nu^2)}{ 1 + \frac{1}{2\nu} \log{\frac{1+\nu}{1-\nu} }}
%

\begin{figure}[h!tb]
	\centering
	\subfigure[]{\includegraphics[width=0.67\linewidth]{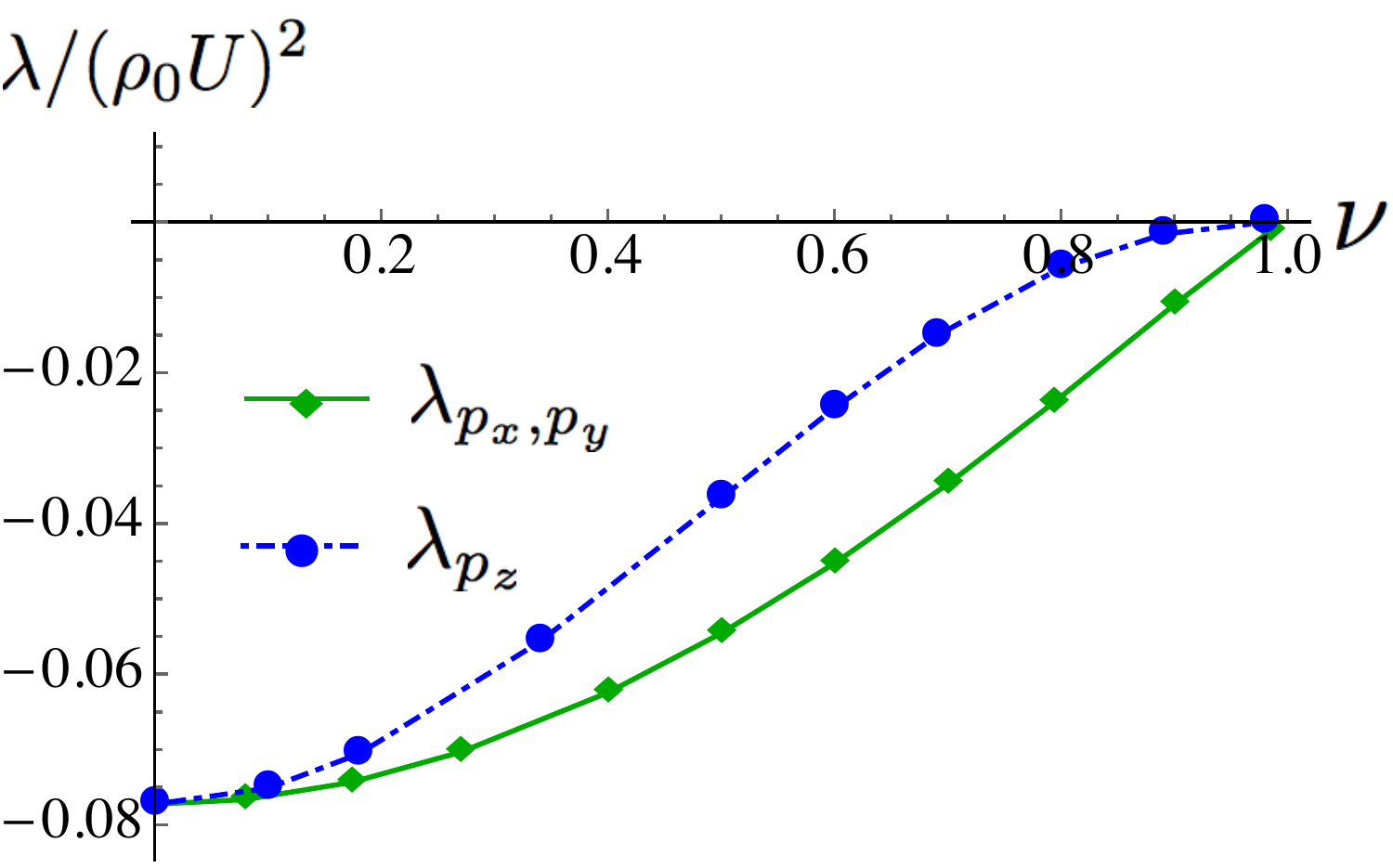}}  \quad \subfigure[]{\includegraphics[width=0.73\linewidth]{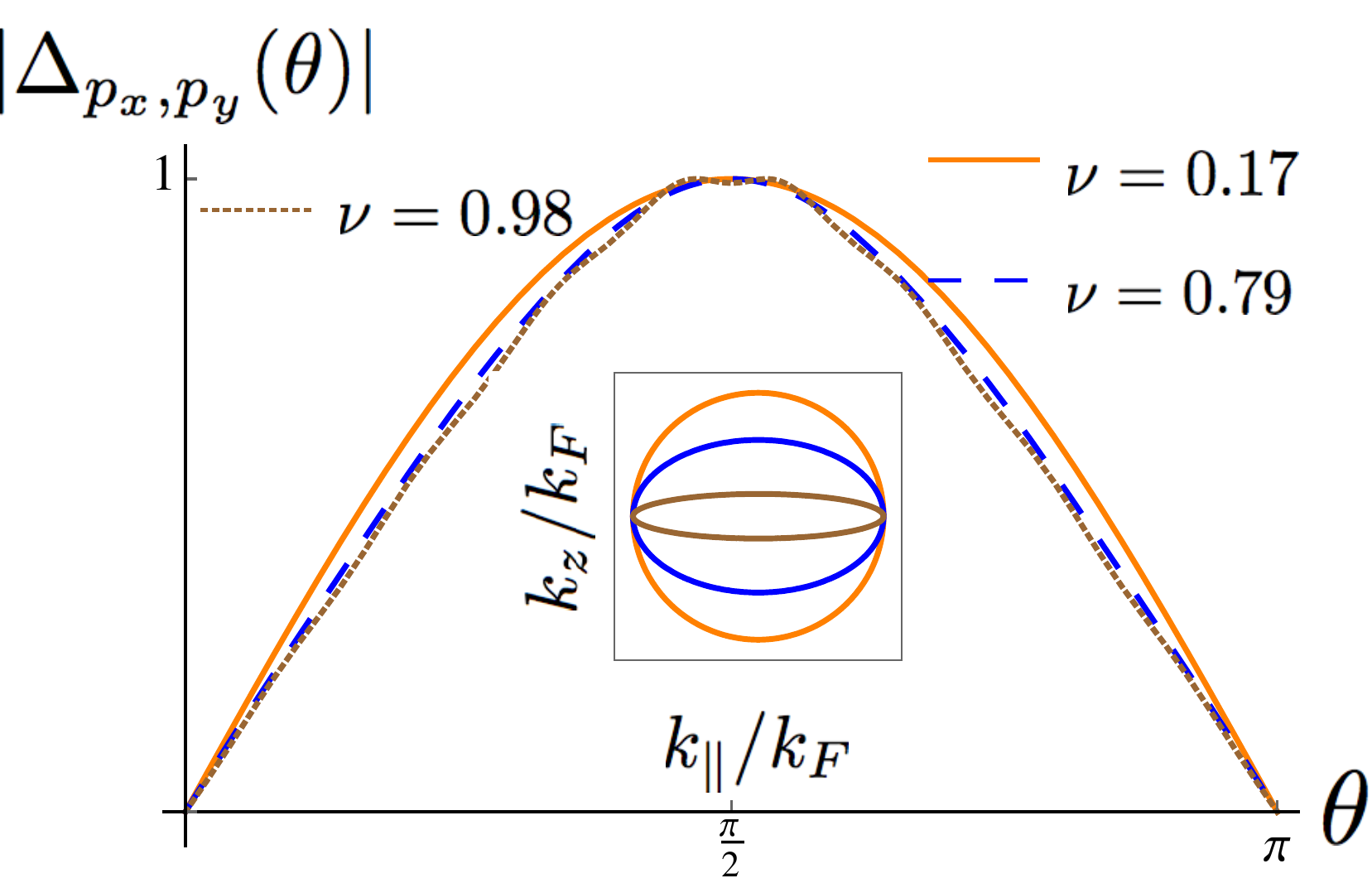} }
\caption{(Color online). (a) Pairing eigenvalues in the oblate regime by expansion in orthogonal polynomials (see Appendix \ref{app:polynomial_expansion}). (b) The ground state order parameter.}
	\label{fig:OblateNumericalSolutions}
\end{figure}

Again solving the integral equation in \eqref{eq:Eigenmodeequation} by expansion in orthogonal polynomials yields the results shown in Fig.\@ \ref{fig:OblateNumericalSolutions}. The order parameters $\Delta_{p_x, p_y}$, with just minor corrections to the $\sin{\theta}$ profile even for large eccentricity, have the highest $T_c$ for an oblate Fermi surface. Within the representation $E_{u}$ a proper linear combination of $p_x$ and $p_y$ is one that minimizes the Ginzburg-Landau free energy that respects the point group symmetries~\cite{GLphaseDiagram}. In our case this is found to be the chiral combinations $p_x \pm i p_y$ for all eccentricities. 

We emphasize that close to $T_c$ the splitting in order parameters is in principle measurable experimentally by applying strain. The $p_z$ order parameter has a horizontal line node whereas the $p_x \pm i p_y$ has point nodes, yielding low-energy density of state power laws $E^2$ and $E^3$, respectively~\cite{SigristUnconSC}. Furthermore, the orders can be distinguished by the heat capacity jump $ \f{ \Delta C }{C_n}\lvert_{T = T_c} = 12 \langle \lvert \tilde{\Delta} \rvert^2 \rangle/(7\zeta(3) \langle \lvert \tilde{\Delta} \rvert^4 \rangle )$, where $\mathrm{max}_{\bo{k}} \tilde{\Delta}_{\bo{k}} = 1$ and the averages are taken over the Fermi surface~\cite{SigristUnconSC}. Using the spherical harmonics $\cos{\theta}$ and $\exp(\pm i \phi) \sin{\theta}$, which are good approximations to the real eigenstates at small eccentricities, we find $ \f{ \Delta C }{C_n}\lvert_{T = T_c}$ to be $20/(21 \zeta(3)) \approx 0.792$ and $10/(7\zeta(3)) \approx 1.19$ with second-order corrections in $\nu$ in the prolate and oblate regime, respectively.

%
%%
%%%
\section{Conclusions}
\label{sec:Conclusion}
%%%
%%
%
The perturbative framework of weak-coupling provided a substantial step forward in the understanding of superconductivity arising from purely repulsive interactions~\cite{KohnLuttinger, Layzer1968, Gorkov60, dwaveinstabHirsch, PairinstabChubukov, PRBRGFlowHubbard}, yet the gap between two and three dimensions previously remained unexplored. In this paper, we have filled this gap with a description of how the order parameter symmetries compete as a function of $t_{\perp}$. Despite studying a model with only nearest-neighbor hopping, we have found the single-band tight-binding model of Eq.~\eqref{eq:Hamiltonian} to contain a rich complexity of $p$- and $d$-wave phases. Our overall observation has been that $d$-wave order tends to win close to half-filling whereas $p$-wave phases win at low filling, with the van Hove lines $\mu^{(x,y)} = -2t_{\perp}$ and $\mu^{(z)} = -4t_{\parallel} + 2t_{\perp}$ dictating the regions of enhanced $T_c$.

Equation \eqref{eq:Hamiltonian} could serve as a basic platform from which to model experimentally-observed systems. To facilitate a match to known systems it would likely be necessary to include further terms in the tight-binding expression, multiple bands, and spin-orbit coupling. For example, applying the weak-coupling scheme to three-dimensional materials such as SrPtAs~\cite{SrPtAsdwave}, FeSe~\cite{KreiselFeSe}, and URu$_2$Si$_2$~\cite{URu2Si2Reveiw} could yield valuable insights. Moreover, the chiral $C = -3$ phase shown in Fig.\@ \ref{fig:OPexamples} (a) displays large relative magnitude differences in the $k_z = 0$ and the $k_z = \pi$ planes for a corrugation of just $t_{\perp} = 0.1$. Since corrugation effects are estimated to be of that order in Sr$_2$RuO$_4$\cite{SrRuOBergmann2003}, this motivates the use of a fully 3D model for future study of this material. Such a 3D model would be especially useful to study the (near-)nodal structure of the superconducting gap (horizontal, vertical, or point nodes), as well as its fate across a van Hove singularity under uniaxial strain \cite{Hicks283, Science2017}.

\begin{acknowledgments}
We thank Fedor \v{S}imkovic for enlightening discussions, and Srinivas Raghu for useful comments. We are grateful to Cathrine Kallin for valuable comments. H.\@ S.\@ R.\@ acknowledges the Aker Scholarship and thanks Yoshiteru Maeno and Daniel Agterberg for useful discussions. F.\@ F.\@ acknowledges support from the Astor Junior Research Fellowship of New College, Oxford. T.\@ S.\@ acknowledges support from the Emergent Phenomena in Quantum Systems initiative of the Gordon and Betty Moore Foundation. S.\@ H.\@ S.\@ is supported by EPSRC Grant No.\@ EP/I031014/1 and  No.\@ EP/N01930X/1.
\end{acknowledgments}

%
%%
%%%
%%%%
%\newpage
\counterwithin{figure}{section}

\begin{appendix}

%
%%
%%%
\section{The weak-coupling approach}
\label{app:MethodDetails}
\begin{figure}[h!tb]
	\centering
\includegraphics[width=0.80\linewidth]{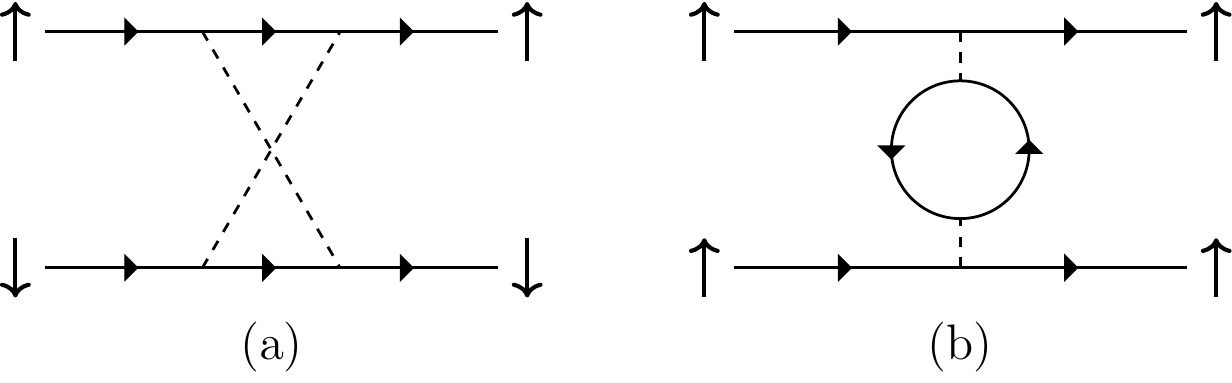}
\caption{The non-trivial diagrams in the (a) singlet and (b) triplet channel for the vertex $\Gamma^{s/t}$ at second order in the interaction $U$. Full lines correspond to the bare propagator, dashed lines the Hubbard interaction, and vertical arrows indicate spin. Incoming states of momenta $\bo{k}$, $-\bo{k}$ are scattered onto outgoing momentum states $\bo{k}'$, $-\bo{k}'$.}
\label{fig:Diagrams}
\end{figure}

%%%
%%
%
The weak-coupling procedure of Ref.\@ \onlinecite{PRBRGFlowHubbard} builds on the seminal work of Refs.\@ \onlinecite{Gorkov60, KohnLuttinger}, and the interaction is treated perturbatively. Rephrasing the main result of Ref.\@ \onlinecite{PRBRGFlowHubbard} the critical temperature permits the weak-coupling expansion $T_c \sim \omega \exp(-\alpha_2 (t/U)^2 -\alpha_1 (t/U) - \alpha_0 ) [1 + \pazocal{O}(U/t) ]$, where $\omega \gg U^2/t$ is the bandwidth. The perturbative results are asymptotically exact in the sense that $\lim_{U \to 0} (U/t)^2 \log(\omega/T_c) = \alpha_2$. Here, $\alpha_2 = (U/t)^2/\lvert \lambda_0 \rvert$, where $\lambda_0$ is the most negative eigenvalue of the two-particle vertex calculated to order $U^2$. The diagrams involved at this order are displayed in Fig.\@ \ref{fig:Diagrams}. The contribution in the triplet channel, shown in Fig.\@ \ref{fig:Diagrams} (b), yields the vertex $\Gamma_{(b)}^t = -\chi(\bo{k}-\bo{k}')$. In the singlet channel, one diagram (Fig.\@ \ref{fig:Diagrams} (a)), again up to second order, contributes to the non-trivial singlet subspace, i.e.\@ where trivial $s$-wave pairing is excluded. Trivial $s$-wave is excluded by the on-site repulsive interaction. The effective vertex in the non-trivial singlet channel is $\Gamma_{(a)}^s = \chi(\bo{k}+\bo{k}')$.

The ground state to leading order in perturbation theory is thus calculated by diagonalization of the matrix $g$, as given in Eq.\@ \eqref{eq:gmatrixtriplet}, and the onset of superconductivity for an order parameter given by the eigenvector (Eq.\@ \eqref{eq:GapFunction}) is identified from the most negative eigenvalue.

The weak-coupling treatment employed here is similar to that of Refs.\@ \onlinecite{PhaseDiagramSimkovic, PhaseDiagramHlubina, Shankar94, Polchinski92} and can be formulated in terms of solving the Bethe-Salpeter equation in the particle-particle channel. At second order, the shape of the Fermi surface is not renormalized. For generic Fermi surface shapes, away from finely tuned points such as perfect nesting or van Hove singularities, this is ensured when the bandwidth is $\omega \gg U^2/t$. In this limit, the susceptibilities that enter the calculation are those of the unperturbed Fermi liquid. Altered effective interactions from renormalization of the Fermi surface thus becomes relevant only at higher order in $U$. 

%
%%
%%%
\section{Fate of the $E_u$ phase}
\label{app:EuFate}
%%%
%%
%
A pocket of the $E_u$ phase appears close to $t_{\perp} = 0$ in the phase diagram.
\begin{figure}[h!tb]
	\centering
\subfigure[]{\includegraphics[width=0.47\linewidth]{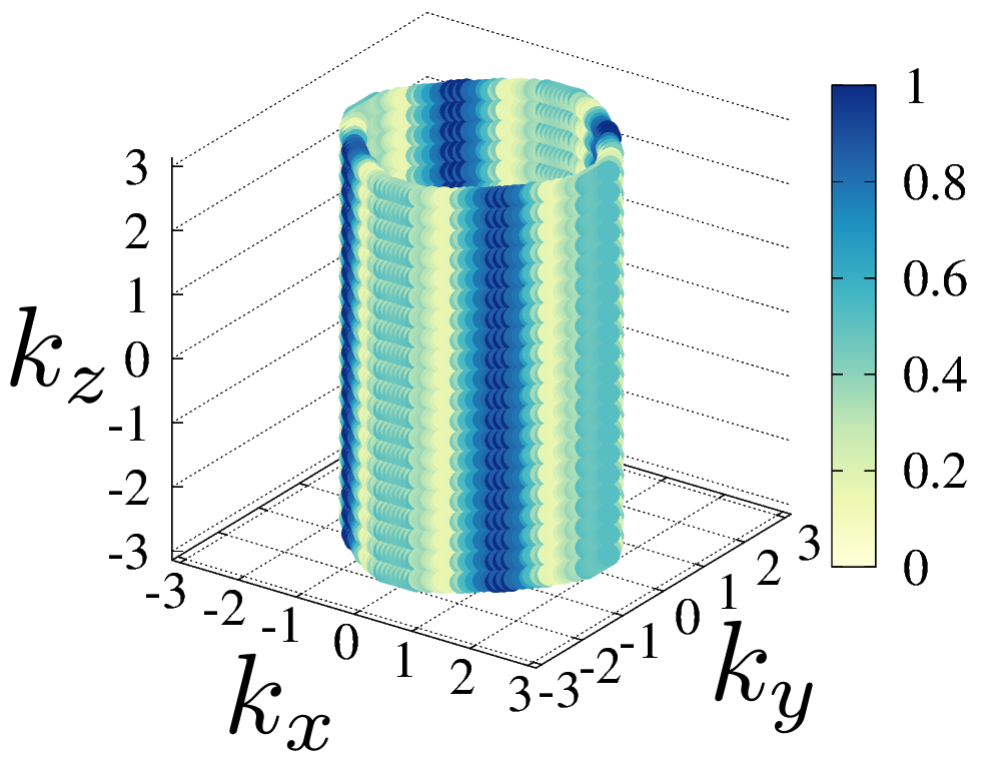}}\quad \subfigure[]
{\includegraphics[width=0.47\linewidth]{pip_mu1c2_tperp0c1.pdf}}\quad \subfigure[]{\includegraphics[width=0.47\linewidth]{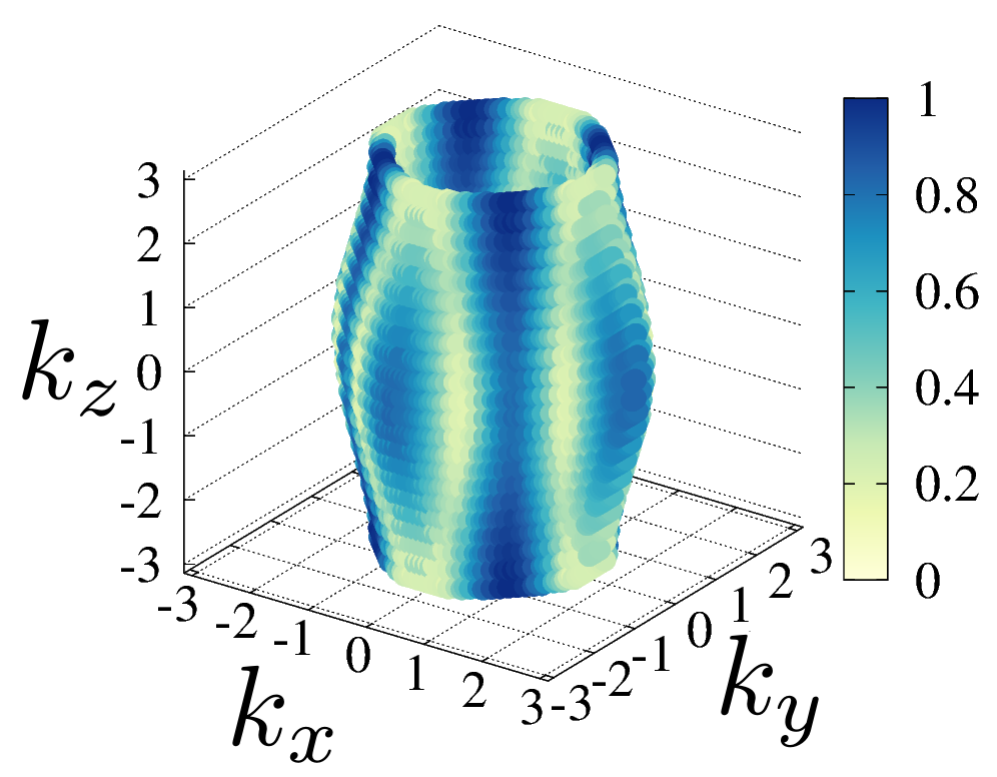}} \quad \subfigure[]{\includegraphics[width=0.47\linewidth]{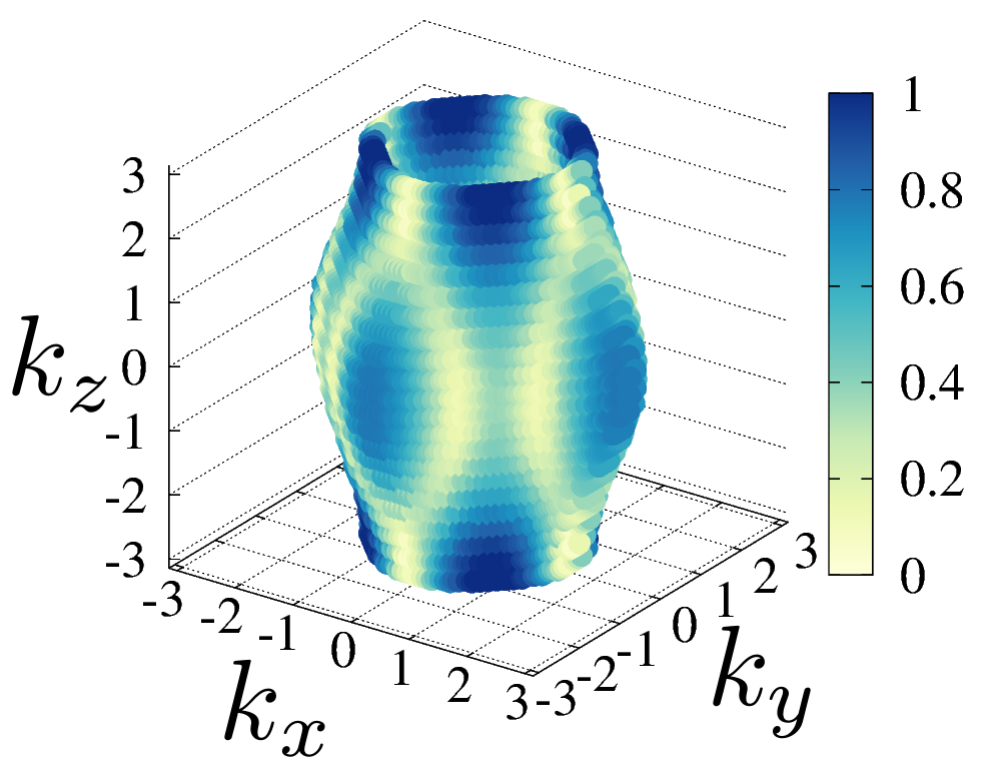}}\quad \subfigure[] {\includegraphics[width=0.75\linewidth]{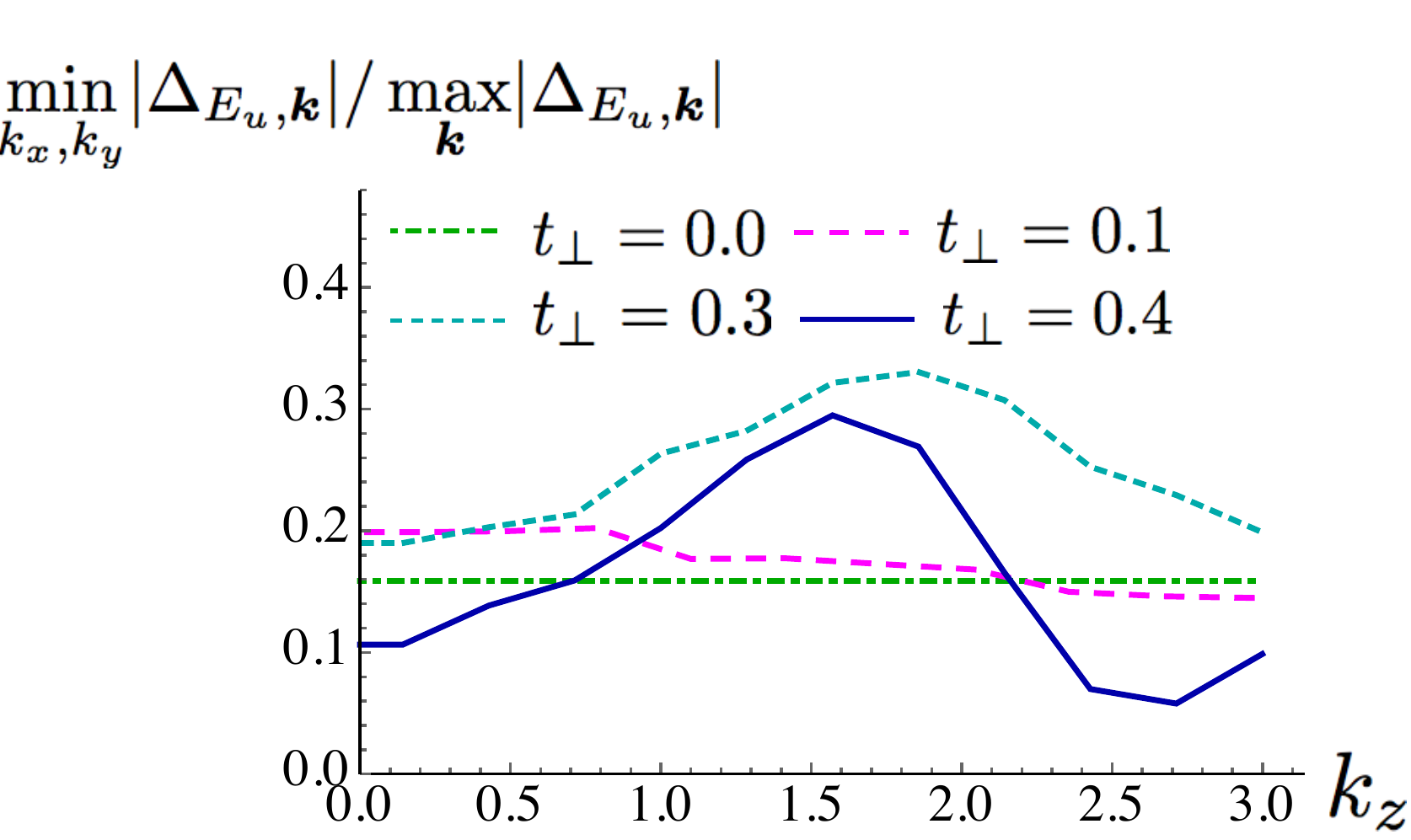}}
\caption{(Color online). The ($E_u$) chiral $p_x + ip_y$ order parameter (magnitude) at $\mu = -1.2$ for $t_{\perp}$ being (a) $0.0$, (b) $0.1$, (c) $0.3$, and (d) $0.4$. Note that the order parameters in (c) and (d) are disfavored compared to $d_{x^2-y^2}$. In (e) we show the gap minima (symmetric in $k_z$) as a function of $k_z$.}
\label{fig:OPpxexampleshighfilling}
\end{figure}
The chiral order parameter, which has a substantial component of the lattice harmonics $\sin(3k_x) + i \sin(3k_y)$, develops an interesting and unfavorable dependency of $k_z$ as $t_{\perp}$ increases. This is shown in Fig.\@ \ref{fig:OPpxexampleshighfilling}. 

%
%%
%%%
\section{Expansion in orthogonal polynomials}
\label{app:polynomial_expansion}
%%%
%%
%
The integral equation can be solved by expanding the integration kernel and its eigenfunctions in appropriate orthogonal polynomials. Given that we seek an eigenfunction of Eq.\@ \eqref{eq:Eigenmodeequation}, we apply the separational ansatzes $\psi_{p_x\pm i p_y}(\theta, \phi) = \tilde{\psi}(\theta) e^{\pm i\phi}$ and $\psi_{p_z}(\theta, \phi) = \tilde{\psi}(\theta)$ in the two sectors of interest, respectively. In either case, when integrating over $\phi'$, this reduces the problem to a one-variable integral equation of the form 
\begin{equation}
\int_{-1}^{1}\D u' \hspace{1mm} K(u \vert u') \tilde{\psi}(u') = \tilde{\lambda}(\nu) \tilde{\psi}(u),
\label{eq:Formulation}
\end{equation}
with $u = \cos{\theta}$ and the kernel being symmetric in $u$, $u'$. We expand the solution in a set of known orthogonal polynomials $p_n(u)$ with appropriate weights $w(u)$ and undetermined coefficients $a_n$. The kernel is expanded in the same set with (a priori undetermined) weights $f_n(u')$:
\begin{align}
\tilde{\psi}(u) &= \sum_n a_n w(u) p_n(u), \label{eq:EigenfunctionExpansion} \\
K(u \vert u') &= \sum_{n} f_n(u') p_n(u).
\label{eq:KernelExpansion}
\end{align}
Inserting the latter expression in Eq.\@ \eqref{eq:Formulation} and comparing with the original formulation shows that the integral equation is reduced to solving the matrix eigenvalue problem
\begin{equation}
\tilde{\lambda} a_n = \sum_m A_{nm} a_m,
\label{eq:MatrixEigenValues}
\end{equation}
where 
\begin{align}
f_m(u') &= c_m \int_{-1}^{1} \D u \hspace{1mm} p_m(u) K(u \vert u'), \label{eq:Weights} \\
A_{nm} &= c_n \int_{-1}^{1} \D u' \int_{-1}^{1} \D u \hspace{1mm} p_n(u) K(u \vert u')  p_m(u'),
\label{eq:MatrixElements}
\end{align}
with $\int_{-1}^1 \D u \hspace{1mm} p_n(u) p_m(u) = c_n \delta_{n,m}$. Assuming that the expansion series converges, one in practice truncates the eigenvalue problem at some finite dimension. Due to the expected form of the solution for low eccentricity, we pick Legendre polynomials, $P_n(u)$ with $w(u) = 1$, and Chebyshev polynomials of the second kind, $U_n(u)$ with $w(u) = \sqrt{1-u^2}$, in the $p_z$ and $p_x\pm ip_y$ sectors, respectively. Solving the matrix eigenvalue problem numerically (truncating the coefficients at $N$ with the maximal tolerance $\lvert a_N/a_0 \rvert \lesssim 10^{-2}$) results in Figs.\@  \ref{fig:ProlateNumericalSolutions} and \ref{fig:OblateNumericalSolutions}.

\end{appendix}

\bibliography{RGSuperconductivityRefs}

\end{document}